# *N*-ALKANOATE + *N*-ALKANE MIXTURES: FOLDING OF HYDROCARBON CHAINS OF *N*-ALKANOATES

Juan Antonio González,* Fernando Hevia, Luis Felipe Sanz, Daniel Lozano-Martín, Isaías García de la Fuente, José Carlos Cobos

G.E.T.E.F., Departamento de Física Aplicada, Facultad de Ciencias, Universidad de Valladolid, Paseo de Belén, 7, 47011 Valladolid, Spain.

*Corresponding author, e-mail: jagl@termo.uva.es

**Abstract**

The mixtures $CH_3(CH_2)_{u-1}COO(CH_2)_{v-1}CH_3$ ($u$ = 5-13, $v$ = 1,2; $u$ = 1,2,3; $v$ =3,4; $u$ =1,2,4, $v$ =5) + $n$-alkane have been investigated on the basis of excess molar functions, enthalpy ($H_m^E$), volume ($V_m^E$), isobaric heat capacity ($C_{pm}^E$), and isochoric internal energy ($U_{Vm}^E$), and viscosity data, and by means of different models (Flory, Grunberg-Nissan and Bloomfield-Dewan). Solutions are characterized by weak orientational effects. Large structural effects are encountered in a number of systems, such as those containing pentane. The variation with the ester size of the difference between the standard enthalpy of vaporization at 298.15 K of an ester and that of the homomorphic alkane along an homologous series formed by methyl or ethyl $n$-alkanoates reveals the existence of structural changes in longer $n$-alkanoates, which lead to stronger interactions between them. A similar result is obtained from values of cohesive energy density. The variation of $V_m^E$ values of the corresponding heptane mixtures supports this statement. The observed decrease of $H_m^E$ for systems with a given $n$-alkane (heptane, e.g.) seems to be more related to the COO group is more sterically hindered than to interactional effects. The $U_{Vm}^E(n)$ function ($n$ is the number of C atoms in the $n$-alkane) shows a minimum for systems with esters characterized by ($u \geq 4$, $v$ =1); ($u \geq 7$, $v$ =2), or ($u \geq 1$, $v$ =4,5). A similar dependence of $U_{Vm}^E(n)$ was encountered for $n$-alkane mixtures involving cyclic molecules (cyclohexane, benzene). This result suggests that certain $n$-alkanoates, in an alkane medium, can form quasi-cyclic structures. Viscosity data are well described by means of free volume effects only. For systems with butyl ethanoate or methyl decanoate, the variation of $\Delta\eta(n)$ (deviation of dynamic viscosity) is consistent with that of $U_{Vm}^E(n)$, which supports the existence of the mentioned cyclic structures in these esters. The Flory model provides poor results on $H_m^E$ for systems characterized by large structural effects. Results are improved when the model is applied to $U_{Vm}^E$ data.

# 1. Introduction

Esters are interesting compounds in view of their many different applications. For example, they are used in food and cosmetics industries where their pleasant odors and antifungal properties are determinant [1,2]. Polyhydroxy alkanoates (PHA) are bio-based polyesters, which have become an alternative for petrochemicals since they can be synthesized by microbial cultures grown on renewable materials in clean environments [3-5]. PHAs have also medical applications [6,7]. On the other hand, biofuels, composed by long methyl or ethyl $n$-alkanoates have gained interest to replace petroleum fuels due to their small production costs, renewability and great potential to reduce harmful gas emissions [8-12].

In the previous work of this series [13], orientational and structural effects (i.e, effects related to different shape, size or free volume of the solution components) in the mixtures linear ester + $n$-alkane were investigated using excess molar functions, enthalpy ($H_m^E$), volume ($V_m^E$), isobaric heat capacity ($C_{pm}^E$) and isochoric internal energy ($U_{Vm}^E$), and by means of the application of the Flory model [14] and the Kirkwood-Buff formalism [15]. However, in the mentioned study only some esters of formula $CH_3(CH_2)_{u-1}COO(CH_2)_{v-1}CH_3$ were considered: methyl alkanoates ($v = 1$; $u = 1$-5); ethyl ($v = 2$), propyl ($v = 3$) or butyl ($v = 4$) alkanoates ($u = 1$-3) and pentyl alkanoates ($v = 5$; $u = 1,2$). The present work extends the research to similar mixtures involving longer $n$-alkanoates: methyl or ethyl alkanoates ($u = 5$-13, $v = 1,2$), or other esters not considered previously, (pentyl pentanoate, e.g.). This is important since in the decade of 1970, starting from some spectroscopic measurements [16,17], Dusart et al., on the basis of $H_m^E$ [18], $V_m^E$ [19] and viscosity [20] data for $CH_3(CH_2)_{u-1}COOCH_3$ ($u = 1$-5) + heptane mixtures, explored the possibility of that some esters (e.g, methyl butanoate) may form quasi-cycles in such systems. The formation of these structures was ascribed to a weak intramolecular interaction between the C=O part of the ester group and an H atom of the terminal $CH_3$ group of the alkyl chain of the $n$-alkanoate. Here, not only $H_m^E$, $V_m^E$ and viscosity data are taken into account, but also special attention is paid to $U_{Vm}^E$. We have demonstrated that this is a key thermodynamic function when investigating cyclic molecule + $n$-alkane mixtures which are characterized by two main features: (i) interactions are mainly of dispersive type (i.e. they arise from dispersion London forces); (ii) solutions with shorter $n$-alkanes show large structural effects [21,22]. Under these conditions, the variation of $U_{Vm}^E$ with $n$, the number of C atoms in the $n$-alkane, depends on two competing contributions. The first leads to decreased values of $U_{Vm}^E$, and is due to a lower ability of longer $n$-alkanes to break interactions between cyclic molecules. The second is endothermic and arises from the breaking of correlations of molecular orientations of longer $n$-alkanes. If the first contribution is dominant, then $U_{Vm}^E$ decreases when *n* is increased. This is the case of mixtures with chlorobenzene, bromobenzene, 1,2,4-trichlorobenzene, or 1-chloronaphthalene [21]. If the second contribution is dominant, $U_{V\text{m}}^{\text{E}}$ increases in line with *n*. This behavior is encountered in systems involving cyclohexane or benzene [22]: $U_{V\text{m}}^{\text{E}}$ decreases up to *n* = 8 and from *n* ≥ 10 increases, which indicates that, for the solutions with longer *n*-alkanes, the second contribution to $U_{V\text{m}}^{\text{E}}$ is dominant. Finally, it must be remarked that, in order to complete our study, the present mixtures are also investigated by means of the following models: Flory [14], Grunberg-Nissan [23] and Bloomfield-Dewan [24].

2. **Models**

*2.1    Flory's model*

The essential hypotheses of the theory [14,25-27] can be found elsewhere [28]. The basic assumption of the theory is that of random mixing. The explicit expression of the Flory equation of state is:

$$\frac{\hat{P}\hat{V}_{\text{m}}}{\hat{T}} = \frac{\hat{V}_{\text{m}}^{1/3}}{\hat{V}_{\text{m}}^{1/3}-1} - \frac{1}{\hat{V}_{\text{m}}\hat{T}} \tag{1}$$

where $\hat{V}_{\text{m}} = V_{\text{m}}/V_{\text{m}}^*$; $\hat{P} = P/P^*$ and $\hat{T} = T/T^*$ stand for the reduced volume, pressure and temperature, respectively ($V_{\text{m}}$ is the molar volume of the mixture). Equation (1) is valid for pure liquids and liquid mixtures. For pure liquids, the reduction parameters, $V_{\text{mi}}^*$, $P_i^*$ and $T_i^*$ can be obtained from density, $\alpha_{pi}$ (isobaric expansion coefficient) and $\kappa_{Ti}$ (isothermal compressibility) data. Expressions for reduction parameters of mixtures can be found elsewhere [28]. $H_{\text{m}}^{\text{E}}$ values are obtained from:

$$H_{\text{m}}^{\text{E}} = \frac{x_1 V_{\text{m1}}^* \theta_2 X_{12}}{\hat{V}_{\text{m}}} + x_1 V_{\text{m1}}^* P_1^* \left(\frac{1}{\hat{V}_{\text{m1}}} - \frac{1}{\hat{V}}\right) + x_2 V_{\text{m2}}^* P_2^* \left(\frac{1}{\hat{V}_{\text{m1}}} - \frac{1}{\hat{V}}\right) \tag{2}$$

where all the symbols have their usual meaning [28]. The reduced volume of the mixture, $\hat{V}_{\text{m}}$, is obtained from the equation of state. Therefore, the molar excess volume can be also calculated:

$$V_{\text{m}}^{\text{E}} = (x_1 V_{\text{m1}}^* + x_2 V_{\text{m2}}^*)(\hat{V}_{\text{m}} - \varphi_1 \hat{V}_{\text{m1}} - \varphi_2 \hat{V}_{\text{m2}}) \tag{3}$$

*2.2 Viscosity models*

*2.2.1 Grunberg-Nissan equation*

This equation, for the correlation of dynamic viscosity ($\eta$) data, is [23]:

$$\ln \eta = x_1 \ln \eta_1 + x_2 \ln \eta_2 + x_1 x_2 G_{12} \qquad (4)$$

where $\eta_i$ stands for the dynamic viscosity of component i, and $G_{12}$ is an adjustable parameter.

*2.2.2 Bloomfield-Dewan's model*

This theory [24] combines the absolute reaction rate model [29] with the free volume theory [30]. The former relates viscosity to the free energy required for a molecule to flow from an equilibrium position to a new one, overcoming the attractive interactions caused by its neighbours. The latter relates viscosity to the probability of occurrence of an empty neighboring site where a molecule can jump. Thus, dynamic viscosity can be determined from the equation:

$$\ln \eta = (x_1 \ln \eta_1 + x_2 \ln \eta_2) + \alpha \ln \eta_{fv} + \beta \ln \eta_{ar} \qquad (5)$$

This means that the probability for viscous flow is calculated as the product of the probabilities of having the sufficient activation energy and of the existence of an empty site [24,31]. The parameters $\alpha, \beta$ are weighting factors with values between 0 and 1. In the present application, calculations were performed assuming that $\beta = 0$, that is, neglecting the residual contribution to $\eta$ and taking into account only free volume effects. In equation (5), $\ln \eta_{fv}$ arises from free volume effects and it is obtained from the expression:

$$\ln \eta_{fv} = \frac{1}{\hat{V}_m - 1} - \frac{x_1}{\hat{V}_{m1} - 1} - \frac{x_2}{\hat{V}_{m2} - 1} \qquad (6)$$

Here, $\hat{V}_m$ and $\hat{V}_{mi}$ are the reduced volumes of the mixture and of component i, respectively, defined as in the Flory model (see above) [28].

### 3  Model calculations and results

Values of physical properties of *n*-alkanoates, required for calculations, are collected in Table S1 (supplementary information; see also Table 1 of reference [13] for other *n*-alkanoates). For most of the *n*-alkanes considered, the values used have been taken from a previous application [32]. For tridecane, pentadecane and heptadecane, the corresponding values are also included in Table 1.

### 3.1 Flory's model

For the $n$-alkanoate + pentane, or + heptane or + pentadecane mixtures, Table 1 lists values of the interaction parameter $X_{12}$ determined from experimental $H_m^E$ data at equimolar composition and 298.15 K following the method provided in reference [33]. Experimental results on $V_m^E$ at the same conditions are also compared with theoretical values in Table 1, which, in addition, contains the relative standard deviations for $H_m^E$ defined as:

$$\sigma_r(H_m^E) = \left[ \frac{1}{N} \sum \left( \frac{H_{m,exp}^E - H_{m,calc}^E}{H_{m,exp}^E} \right)^2 \right]^{1/2} \quad (7)$$

where $N$ is the number of experimental data points. For the sake of completeness, Table S2 collects results for systems including other $n$-alkanes. A comparison between experimental $H_m^E$ results and model calculations is shown in Figures 1-2 and S1.

**Table 1.** Excess molar enthalpies, $H_m^E$, volumes, $V_m^E$, and isochoric internal energies, $U_{Vm}^E$, at equimolar composition and 298.15 K, for $n$-alkanoate (1) + $n$-alkane (2) mixtures. Values of the Flory interaction parameter, $X_{12}$, and of the relative standard deviations for $H_m^E$ and $U_{Vm}^E$ ($\sigma_r(H_m^E)$ and $\sigma_r(U_{Vm}^E)$; see equation 7), are also included.

| $n^a$ | $N^b$ | $H_m^E$ / J mol$^{-1}$ | $X_{12}$ / J cm$^{-3}$ | $\sigma_r(H_m^E)$ | Ref. | $V_m^E$ / cm$^3$mol$^{-1}$ Exp. | $V_m^E$ / cm$^3$mol$^{-1}$ Flory | Ref. | $U_{Vm}^E$ / J mol$^{-1}$ | $X_{12}$ / J cm$^{-3}$ | $\sigma_r(U_{Vm}^E)$ $^c$ |
|---|---|---|---|---|---|---|---|---|---|---|---|
| methyl hexanoate + $n$-alkane | | | | | | | | | | | |
| 5 | 21 | 721 | 27.50 | 0.189 | 57 | − 0.465 | − 0.257 | 59 | 842 | 31.86 | 0.128 |
| 7 | 18 | 838 | 27.69 | 0.058 | 34 | 0.278 | 0.463 | 59 | 757 | 25.04 | 0.054 |
| 15 | 18 | 1220 | 33.27 | 0.055 | 73 | 0.814 | 1.032 | 74 | 956 | 26.14 | 0.065 |
| methyl heptanoate + $n$-alkane | | | | | | | | | | | |
| 5 | 20 | 631 | 23.00 | 0.193 | 57 | − 0.639 | − 0.540 | 58 | 800 | 28.69 | 0.117 |
| 7 | 17 | 741 | 22.58 | 0.080 | 34 | − 0.082 | − 0.117 | 58 | 718 | 22.24 | 0.081 |
| 15 | 17 | 1101 | 27.59 | 0.062 | 73 | 0.701 | 0.924 | 74 | 874 | 21.94 | 0.073 |
| methyl octanoate + $n$-alkane | | | | | | | | | | | |
| 5 | 19 | 561 | 19.76 | 0.189 | 57 | − 0.826 | − 0.689 | 59 | 778 | 26.64 | 0.114 |
| 7 | 17 | 645 | 18.93 | 0.089 | 34 | − 0.023 | 0.090 | 59 | 652 | 19.13 | 0.060 |
| 15 | 16 | 1001 | 23.26 | 0.064 | 73 | 0.608 | 0.787 | 74 | 806 | 18.72 | 0.073 |
| methyl decanoate + $n$-alkane | | | | | | | | | | | |
| 5 | 19 | 449 | 15.14 | 0.269 | 57 | − 1.070 | − 1.005 | 59 | 739 | 23.45 | 0.139 |
| 7 | 20 | 550 | 14.81 | 0.106 | 34 | − 0.209 | − 0.154 | 59 | 612 | 16.40 | 0.087 |
| 15 | 18 | 831 | 17.00 | 0.061 | 73 | 0.460 | 0.596 | 74 | 683 | 13.97 | 0.063 |
| methyl dodecanoate + $n$-alkane | | | | | | | | | | | |
| 5 | 17 | 357 | 11.92 | 0.348 | 35 | − 1.251 | − 1.139 | 59 | 700 | 20.95 | 0.159 |
| 7 | 17 | 442 | 11.26 | 0.096 | 35 | − 0.325 | − 0.436 | 59 | 537 | 13.49 | 0.055 |
| 15 | 16 | 707 | 13.04 | 0.077 | 73 | 0.354 | 0.435 | 74 | 589 | 10.87 | 0.084 |
| methyl tetradecanoate + $n$-alkane | | | | | | | | | | | |
| 5 | 18 | 305 | 10.11 | 0.354 | 35 | − 1.450 | − 1.609 | 59 | 709 | 20.02 | 0.139 |
| 7 | 18 | 373 | 9.12 | 0.118 | 35 | − 0.467 | − 0.670 | 59 | 516 | 12.23 | 0.082 |
| 15 | 18 | 597 | 10.12 | 0.085 | 73 | 0.273 | 0.286 | 74 | 506 | 8.58 | 0.092 |

| system | $n^a$ | $N^b$ | | | | | | | | | |
|---|---|---|---|---|---|---|---|---|---|---|---|
| ethyl hexanoate + *n*-alkane | | | | | | | | | | | |
| | 5 | 18 | 406 | 14.97 | 0.195 | 45 | − 0.532 | − 0.514 | 45 | 549 | 19.77 | 0.103 |
| | 7 | 17 | 655 | 20.08 | 0.084 | 45 | 0.222 | 0.341 | 45 | 592 | 18.17 | 0.080 |
| | 15 | 17 | 1086 | 27.33 | 0.063 | 45 | 0.743 | 0.942 | 45 | 854 | 21.57 | 0.079 |
| ethyl heptanoate + *n*-alkane | | | | | | | | | | | |
| | 5 | 17 | 302 | 11.21 | 0.232 | 45 | − 0.757 | − 0.745 | 45 | 506 | 17.65 | 0.115 |
| | 7 | 17 | 530 | 15.43 | 0.070 | 45 | 0.091 | 0.122 | 45 | 510 | 14.86 | 0.043 |
| | 15 | 17 | 944 | 21.95 | 0.095 | 45 | 0.645 | 0.774 | 45 | 739 | 17.36 | 0.113 |
| ethyl octanoate + *n*-alkane | | | | | | | | | | | |
| | 5 | 19 | 234 | 8.77 | 0.258 | 46 | − 0.958 | − 0.914 | 46 | 491 | 16.38 | 0.129 |
| | 7 | 17 | 437 | 12.16 | 0.041 | 46 | − 0.029 | − 0.028 | 46 | 445 | 12.37 | 0.013 |
| | 15 | 17 | 783 | 17.07 | 0.129 | 46 | 0.572 | 0.654 | 46 | 599 | 13.09 | 0.141 |
| ethyl decanoate + *n*-alkane | | | | | | | | | | | |
| | 5 | 17 | 151 | 6.19 | 0.316 | 46 | − 1.200 | − 1.174 | 46 | 482 | 15.24 | 0.118 |
| | 7 | 19 | 300 | 7.92 | 0.035 | 46 | − 0.237 | − 0.284 | 46 | 370 | 9.63 | 0.034 |
| | 15 | 17 | 582 | 11.29 | 0.106 | 46 | 0.435 | 0.449 | 46 | 444 | 8.62 | 0.126 |
| ethyl dodecanoate + *n*-alkane | | | | | | | | | | | |
| | 5 | 13 | 99 | 4.65 | 0.336 | 46 | − 1.332 | − 1.330 | 46 | 467 | 13.94 | 0.084 |
| | 7 | 18 | 251 | 6.31 | 0.115 | 46 | − 0.403 | − 0.430 | 46 | 375 | 9.10 | 0.089 |
| | 15 | 18 | 437 | 7.72 | 0.132 | 46 | 0.314 | 0.331 | 46 | 333 | 5.89 | 0.155 |
| ethyl tetradecanoate + *n*-alkane | | | | | | | | | | | |
| | 5 | 10 | 74 | 4.09 | 0.639 | 46 | − 1.472 | − 1.500 | 46 | 482 | 14.07 | 0.075 |
| | 7 | 17 | 231 | 5.79 | 0.224 | 46 | − 0.477 | − 0.540 | 46 | 375 | 8.92 | 0.217 |
| | 15 | 18 | 377 | 6.37 | 0.126 | 46 | 0.245 | 0.246 | 46 | 297 | 5.02 | 0.150 |
| propyl ethanoate + *n*-alkane | | | | | | | | | | | |
| | 5 | 17 | 1024 | 44.16 | 0.257 | 75 | 0.131 | 0.544 | 75 | 990 | 42.72 | 0.196 |
| | 7 | 17 | 1199 | 46.95 | 0.032 | 75 | 0.787 | 1.170 | 75 | 969 | 37.96 | 0.076 |
| | 15 | 17 | 1718 | 58.18 | 0.074 | 75 | 1.167 | 1.628 | 75 | 1344 | 46.17 | 0.097 |
| propyl propanoate + *n*-alkane | | | | | | | | | | | |
| | 5 | 17 | 670 | 26.96 | 0.249 | 75 | − 0.168 | − 0.263 | 75 | 714 | 28.55 | 0.220 |
| | 7 | 17 | 893 | 31.76 | 0.091 | 75 | 0.562 | 0.668 | 75 | 736 | 26.19 | 0.082 |
| | 15 | 17 | 1390 | 42.21 | 0.070 | 75 | 0.981 | 1.564 | 75 | 1078 | 33.12 | 0.129 |
| propyl butanoate + *n*-alkane | | | | | | | | | | | |
| | 5 | 19 | 494 | 18.77 | 0.354 | 75 | − 0.379 | − 0.336 | 75 | 595 | 22.35 | 0.267 |
| | 7 | 16 | 747 | 24.53 | 0.091 | 75 | 0.393 | 0.058 | 75 | 633 | 20.81 | 0.112 |
| | 15 | 17 | 1192 | 32.76 | 0.046 | 75 | 0.865 | 1.097 | 75 | 915 | 25.49 | 0.064 |
| butyl ethanoate + *n*-alkane | | | | | | | | | | | |
| | 5 | 18 | 789 | 31.65 | 0.157 | 55 | − 0.224 | 0.064 | 55 | 848 | 34.03 | 0.135 |
| | 7 | 17 | 980 | 34.81 | 0.061 | 55 | 0.548 | 0.782 | 55 | 821 | 29.08 | 0.103 |
| | 15 | 18 | 1447 | 43.49 | 0.075 | 55 | 0.996 | 1.283 | 55 | 1129 | 34.22 | 0.087 |
| butyl propanoate + *n*-alkane | | | | | | | | | | | |
| | 5 | 15 | 577 | 21.91 | 0.193 | 55 | − 0.484 | − 0.290 | 55 | 706 | 26.61 | 0.098 |
| | 7 | 18 | 758 | 24.89 | 0.100 | 55 | 0.359 | 0.492 | 55 | 654 | 21.52 | 0.102 |
| | 15 | 18 | 1208 | 33.13 | 0.082 | 55 | 0.861 | 1.104 | 55 | 932 | 25.66 | 0.101 |
| butyl butanoate + *n*-alkane | | | | | | | | | | | |
| | 5 | 15 | 400 | 15.10 | 0.320 | 55 | − 0.632 | − 0.690 | 55 | 568 | 20.58 | 0.167 |
| | 7 | 17 | 565 | 17.43 | 0.159 | 55 | 0.190 | 0.131 | 55 | 510 | 15.95 | 0.151 |
| | 15 | 17 | 1038 | 25.79 | 0.087 | 55 | 0.761 | 0.856 | 55 | 794 | 19.77 | 0.087 |
| pentyl ethanoate + *n*-alkane | | | | | | | | | | | |
| | 5 | 16 | 636 | 24.21 | 0.217 | 56 | − 0.453 | 0.211 | 56 | 758 | 28.54 | 0.117 |
| | 7 | 17 | 828 | 27.20 | 0.086 | 56 | 0.350 | 0.532 | 56 | 729 | 23.96 | 0.162 |
| | 15 | 20 | 1305 | 35.43 | 0.081 | 56 | 0.882 | 1.068 | 56 | 1019 | 27.69 | 0.089 |
| pentyl propanoate + *n*-alkane | | | | | | | | | | | |
| | 5 | 16 | 468 | 17.19 | 0.104 | 56 | − 0.631 | − 0.559 | 56 | 635 | 22.79 | 0.027 |
| | 7 | 17 | 647 | 19.89 | 0.078 | 56 | 0.156 | 0.265 | 56 | 601 | 18.43 | 0.071 |
| | 15 | 16 | 1056 | 26.51 | 0.071 | 56 | 0.764 | 0.930 | 56 | 809 | 20.46 | 0.090 |
| pentyl pentanoate + *n*-alkane | | | | | | | | | | | |
| | 5 | 15 | 323 | 11.43 | 0.170 | 56 | − 1.026 | − 0.864 | 56 | 601 | 19.75 | 0.076 |
| | 7 | 15 | 448 | 12.44 | 0.114 | 56 | − 0.050 | 0.327 | 56 | 463 | 12.85 | 0.092 |
| | 15 | 19 | 771 | 16.79 | 0.087 | 56 | 0.612 | 0.652 | 56 | 575 | 12.55 | 0.138 |

[a]number of C atoms in the *n*-alkane; [b]number of data points

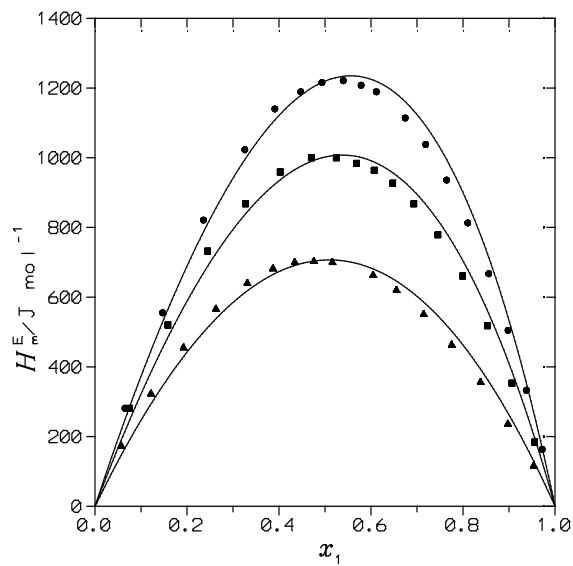

**Figure 1**. $H_\mathrm{m}^\mathrm{E}$ for CH3(CH2)$_{u\text{-}1}$COOCH3 (1) + pentadecane (2) systems at 298.15 K. Symbols, experimental results (for source of data, see Table 1): (●), $u = 5$; (■), $u = 7$; (▲); $u = 11$. Lines, Flory results using interaction parameters listed in Table 1.

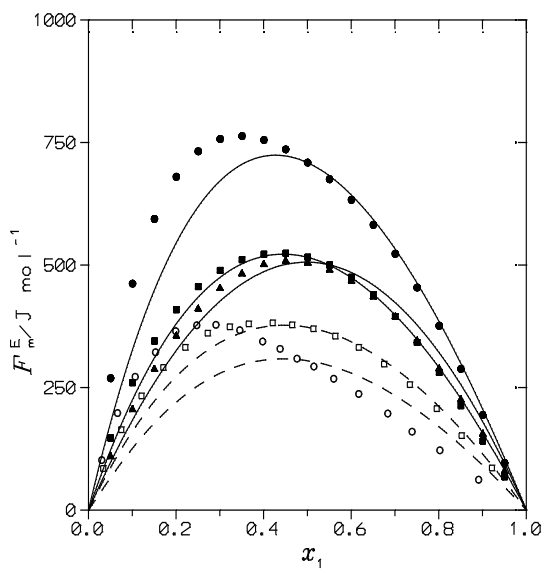

**Figure 2**. Excess molar functions for CH3(CH2)12COOCH3 (1) + *n*-alkane mixtures (2) systems at 298.15 K. Open symbols, $F_\mathrm{m}^\mathrm{E} = H_\mathrm{m}^\mathrm{E}$ (see Table 1 for source of data): (O), pentane; (□), heptane. Full symbols, $F_\mathrm{m}^\mathrm{E} = U_{V\mathrm{m}}^\mathrm{E}$ (this work): (●), pentane; (■), heptane; (▲); pentadecane. Lines, Flory calculations using interaction parameters listed in Table 1: solid lines, $U_{V\mathrm{m}}^\mathrm{E}$; dashed lines, $H_\mathrm{m}^\mathrm{E}$.

*3.2 Viscosity models*

Results from the application of the two viscosity models considered are collected in Table 2, which lists values of the relative deviations of dynamic viscosities, $\sigma_\mathrm{r}(\eta)$, defined similarly to equation (7) (see also Figure 3).

**Table 2.** Dynamic viscosities ($\eta$), deviations in absolute viscosity ($\Delta\eta$), excess molar volumes, $V_m^E$, and deviations of molar Gibbs energy of activation, $\Delta(\Delta G_m^*)$ (equation 11), for $CH_3(CH_2)_{u-1}COO(CH_2)_{v-1}CH_3$ (1) + n-alkane (2) mixtures at 298.15 K, and equimolar composition. Results from the application of the Bloomfield-Dewan's model with $\beta = 0$ (see equation 5) and from the correlation of data using the Grunberg-Nissan's equation (equation 4) are also included.

| (u,v) | n-$C_n$ | $N^a$ | $\eta$ / mPa s | $\Delta\eta$ / mPa s | $\sigma_r(\eta)^b$ | Ref. $\eta$ | $V_m^E$ / cm³ mol⁻¹ | Ref. $V_m^E$ | $G_{12}^c$ | $\sigma_r(\eta)^b$ | $\Delta(\Delta G_m^*)$ / J mol⁻¹ |
|---|---|---|---|---|---|---|---|---|---|---|---|
| (1,1) | 7 | 11 | 0.343 | − 0.034 | 0.023 | 65 | 1.378 | 37 | − 0.376 | 0.006 | − 88 |
| (2,1) | 7 | 12 | 0.378 | − 0.035 | 0.014 | 20 | 0.973 | 37 | − 0.351 | 0.014 | − 144 |
| (3,1) | 7 | 9 | 0.416 | − 0.048 | 0.002 | 65 | 0.723 | 37 | − 0.399 | 0.008 | − 224 |
| (4,1) | 7 | 11 | 0.474 | − 0.051 | 0.010 | 20 | 0.489 | 37 | − 0.276 | 0.009 | − 158 |
| (5,1) | 7 | 9 | 0.539 | − 0.072 | 0.011 | 65 | 0.295 | 76 | − 0.218 | 0.005 | − 130 |
| (7,1) | 7 | 20 | 0.699 | − 0.133 | 0.015 | 65 | − 0.023 | 59 | − 0.032 | 0.005 | − 6 |
| (9,1) | 7 | 9 | 0.918 | − 0.247 | 0.011 | 65 | − 0.209 | 59 | 0.220 | 0.008 | 177 |
| (1,1) | 16 | 9 | 1.451 | − 0.276 | 0.108 | 77 | 1.695 | 77 | 1.178 | 0.050 | 1247 |
| (1,4) | 16 | 12 | 1.610 | − 0.266 | 0.050 | 77 | 1.064 | 77 | 0.434 | 0.016 | 473 |
| (9,1) | 16 | 9 | 2.338 | − 0.147 | 0.009 | 78 | 0.451 | 78 | − 0.136 | 0.003 | − 50 |
| (11,1) | 16 | 9 | 2.776 | − 0.142 | 0.016 | 78 | 0.358 | 78 | − 0.195 | 0.002 | − 109 |
| (13,1) | 16 | 9 | 3.321 | − 0.166 | 0.013 | 78 | 0.285 | 78 | − 0.165 | 0.002 | − 99 |
| (2,3) | 6 | 12 | 0.407 | − 0.059 | 0.022 | 79 | 0.341 | 79 | − 0.237 | 0.013 | − 141 |
| (2,3) | 7 | 9 | 0.476 | − 0.039 | 0.019 | 80 | 0.573 | 80 | − 0.197 | 0.018 | − 108 |
| (2,3) | 8 | 9 | 0.531 | − 0.046 | 0.008 | 80 | 0.712 | 80 | − 0.311 | 0.005 | − 168 |
| (2,3) | 10 | 11 | 0.697 | − 0.054 | 0.005 | 81 | 0.809 | 81 | − 0.255 | 0.003 | − 99 |
| (2,3) | 14 | 10 | 1.185 | − 0.175 | 0.027 | 81 | 0.941 | 81 | − 0.104 | 0.003 | 216 |

$^a$N, number of data points; $^b \sigma_r(\eta) = \left[\frac{1}{N}\sum\left(\frac{\eta_{exp}-\eta_{calc}}{\eta_{exp}}\right)^2\right]^{1/2}$; $^c$adjustable parameter of the Grunberg-Nissan's equation (equation 4).

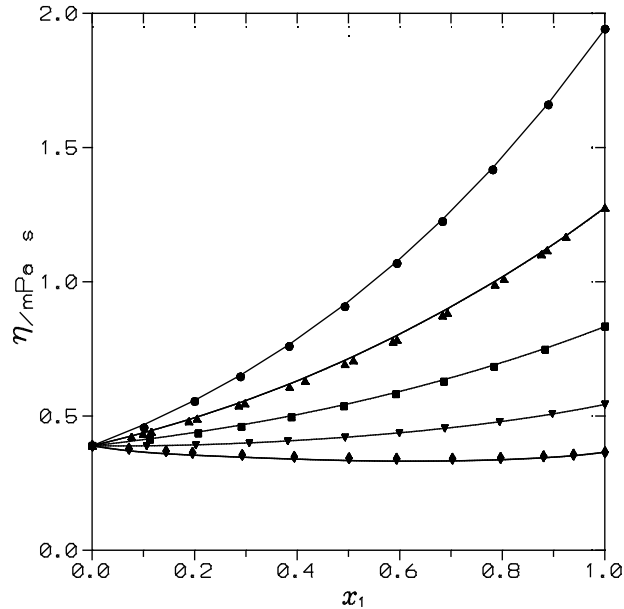

**Figure 3.** Viscosity, $\eta$, of $CH_3(CH_2)_{u-1}COOCH_3$ (1) + heptane (2) systems at 298.15 K. Points, experimental values (for source of data, see Table 2): (♦), $u$ =1, (▼), $u$ = 3; (■), $u$ = 5; (▲), $u$ = 7; (●), $u$ = 9. Lines, results from the application of the Bloomfield-Dewan's model assuming β = 0.

## 4. Discussion

Below, values of the reported thermodynamic properties are given at equimolar composition and 298.15 K.

### 4.1  $CH_3(CH_2)_{u-1}COO(CH_2)_{v-1}CH_3$ (1) + heptane (2)

#### 4.1.1  Excess molar enthalpies

Firstly, mixtures including methyl ($v =1$) or ethyl ($v = 2$) $n$-alkanoates are considered, since the corresponding available database is larger. These systems show positive values of $H_m^E$, which indicates that interactions between like molecules (i.e., interactions between molecules of the same type within the system) are dominant. If one assumes that $u$ increases when $v$ is constant, we note that $H_m^E$ decreases (Tables 1, S2 and Figure 4). For example, $H_m^E$ ($v = 1$)/J mol$^{-1}$ = 1787 $(u =1)$ [34]; 1424 $(u = 2)$ [34], 968 $(u = 4)$ [34], 645 $(u = 7)$ [34], 550 $(u = 9)$ [34]; 442 $(u = 11)$ [35]. On the other hand, it is remarkable that the observed decrease of $H_m^E$ ($v = 1$) is sharper for systems with ($u \leq 4$) and smoother for solutions with ($u \geq 6$). Solutions with $v = 2$ behave similarly (Figure 4). This reveals that dipolar interactions are relatively more important in mixtures including shorter $n$-alkanoates and that such interactions become weaker when the ester size increases [13]. Note that mixtures with methyl ethanoate show miscibility gaps at relatively high temperatures (the upper critical temperature solution of the system with octane is 241.7 K [36]).

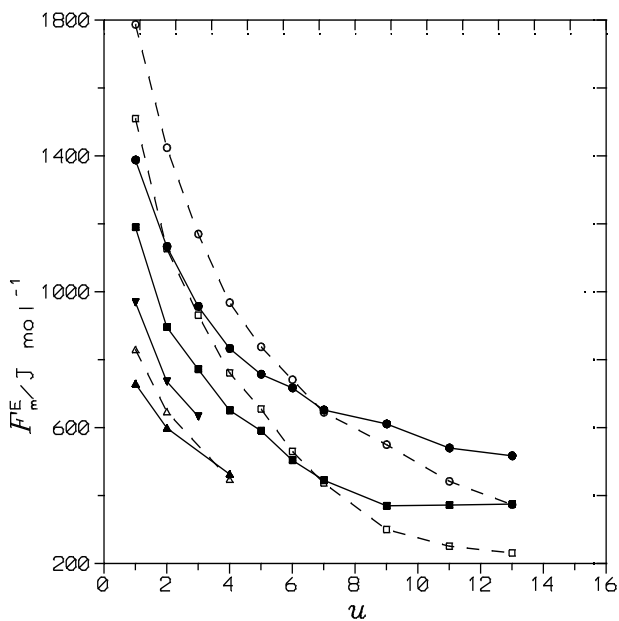

**Figure 4**. Excess molar functions for $CH_3(CH_2)_{u-1}COO(CH_2)_{v-1}CH_3$ (1) + heptane (2) systems at equimolar composition and 298.15 K. Open symbols (joined by dashed lines), $F_m^E = H_m^E$: (O), $v =1$; (□), $v = 2$; (Δ), $v = 5$. Full symbols (joined by solid lines), $F_m^E = U_{Vm}^E$: (●), $v = 1$; (■), $v = 2$; (▼), $v = 3$; (▲); $v = 5$. Lines are for the aid of the eye. For source of data, see Tables 1, S2, S5 and reference [13].

Accordingly with this behaviour, the system methyl ethanoate + heptane shows a $C_{pm}^{E}(x_1)$ curve with two negative minima at the extremes of the concentration range and positive values at its central part (0.91 J mol$^{-1}$ K$^{-1}$) [37], which is a typical feature of mixtures at temperature not far from the UCST [37]. For systems containing longer methyl $n$-alkanoates, the ($C_{pm}^{E}$/J mol$^{-1}$ K$^{-1}$) results become more negative when the ester size increases: $-0.29$ (methyl propanoate); $-0.75$ (methyl octanoate) [38], and $-0.50$ for the mixture with propyl butanoate [39]. Similar trends are encountered in $n$-alkanoate + cyclohexane systems [40]. Note that negative values of $C_{pm}^{E}$, normally ascribed to order destruction, are encountered in systems characterized by dispersive interactions [41,42]. In addition, it should be also taken into account that the COO group becomes more sterically hindered when $u$ increases ($v$ =1,2), an effect that leads to decreased $H_m^E$ values since entails a decrease in the number of ester-ester interactions in comparison to that of pure alkanoates [13,43]. The weakening of the ester-ester interactions may be also investigated by means of the partial excess molar enthalpies at infinite dilution of the $n$-alkanoate, $H_{m1}^{E,\infty}$ (determined from $H_m^E$ measurements over the whole concentration range, Table S3). For mixtures with methyl esters ($v$ = 1), this magnitude decreases up to $u$ = 9, and it is constant for $u$ =11,13 (Table S3): $H_{m1}^{E,\infty}(v=1)$/kJ mol$^{-1}$ = 8.3 ($u$ =1) [44]; 5.0 ($u$ = 3) [34]; 4.2 ($u$ = 5) [34]; 3.5 ($u$ = 7) [34]; 2.7 ($u$ = 11, 13) [35]. In the case of systems containing ethyl $n$-alkanoates ($v$ = 2), the dependence of $H_{m1}^{E,\infty}$ with $u$ is similar: $H_{m1}^{E,\infty}(v=2)$/kJ mol$^{-1}$ = 7.0 ($u$ =1) [45]; 4.5 ($u$ =3) [45] ; 2.7 ($u$ = 6) [45]; 2.0 ($u$ =7) [46]; 1.5 ($u$ = 9, 11) [46]; and 1.8 ($u$ =13) [46]. In order to continue investigating the strength of the interactions between ester molecules, two magnitudes are now considered: $\Delta\Delta H_{vap,i}$ and $D_{ce,i}$. The former is defined as the difference between the standard enthalpy of vaporization at 298.15 K, $\Delta H_{vap,i}$, of a compound characterized by a polar group X (here, X = COO) and that of the homomorphic alkane [47]. The magnitude $\Delta\Delta H_{vap,i}$ is useful to compare the relative changes in intermolecular forces of homomorphic compounds upon replacing a CH$_2$ group by a given X group [47]. $D_{ce,i}$ is the density of cohesive energy determined from the equation [48]:

$$D_{cei} = \frac{\Delta H_{vap,i} - RT}{V_{mi}} \tag{8}$$

and describes the total strength of the solvent structure [48]. Values of $\Delta H_{vap,i}$ for the calculation of the magnitude $\Delta\Delta H_{vap,i}$ have been taken from [49-52] (see Table S4). Some results follow: $\Delta\Delta H_{vap,i}$ ($v$ = 1)/kJ mol$^{-1}$ = 12.8 ($u$ =3); 11.5 ($u$ =5); 10.0 ($u$ = 7); 10.3 ($u$ =9); 10.7 ($u$ =11); 10.9 ($u$ =13); and $\Delta\Delta H_{vap,i}$ ($v$ = 2)/kJ mol$^{-1}$ = 11.1 ($u$ =3); 10.2 ($u$ =5); 8.1 ($u$ =7); 9.0 ($u$ = 9); 8.7 ($u$ =11). We note that $\Delta\Delta H_{vap,i}$ shows a minimum at $u$ = 7, and then slightly increases, remaining roughly constant. This is an unexpected finding, and suggests the existence of some structural change in longer $n$-alkanoates which seems lead to stronger interactions between them. This effect is more pronounced for ethyl $n$-alkanoates. Values of $D_{ce,i}$ (Table S4) support this statement. Thus, $D_{cei}$ ($v$ = 1)/kJ mol$^{-1}$ = 321.6 ($u$ = 3); 307.7 ($u$ = 5); 297.3 ($u$ = 7); 299.6 ($u$ = 9); 301.4 ($u$ = 11); 300.8 ($u$ = 13), and $D_{ce}$ ($v$ = 2)/kJ mol$^{-1}$ = 302.2 ($u$ = 3); 295.8 ($u$ = 5); 285.4 ($u$ = 7); 292.0 ($u$ = 9); 291.3 ($u$ = 11). If one also takes into account the corresponding $H_{m1}^{E,\infty}$ values (see above), the observed decrease of $H_m^E$ for systems with long chain esters seems to be more related to the COO group is more sterically hindered than to interactional effects. It is interesting to note that a similar variation of $\Delta\Delta H_{vap,i}$ and $D_{ce,i}$ is encountered for 2-alkanones (Table S4) and, therefore, it seems that a similar structural change may exist in longer 2-alkanones. It is pertinent to underline that dialkyl ethers behave "normally" and the magnitudes under consideration decreases when the size of the oxaalkane increases (Table S4). The negative values of $\Delta\Delta H_{vap,i}$ obtained for dialkylethers longer than dipropylether reveal that dispersive interactions in such oxaalkanes are dominant by far [47].

Next, we pay attention to the internal pressure ($P_{int,i}$), and, particularly to the ratio $q_i = \dfrac{P_{int,i}}{D_{ce,i}}$. The internal pressure is mainly determined by dispersive and weak dipolar interactions [48,53]. Values for $n$-alkanoates are listed Table S4. They were obtained using the expression:

$$P_{int,i} = \left[\left(\dfrac{\partial U}{\partial V}\right)_T\right]_i = T\dfrac{\alpha_{p,i}}{\kappa_{T,i}} - p \qquad (9)$$

With regards to $q_i$, for associated liquids, $q_i \leq 0.8$, i.e., $P_{int,i} \leq D_{ce,i}$ [53]. Values of $q_i$ ranged between [1.0,1.2] are typical of hydrocarbons (Table S4) and of non-polar liquids [53] (see results for $n$-alkanes and oxaalkanes, Table S4). For $n$-alkanoates, their $q_i$ values are encountered in the mentioned interval (Table S4), and one can assume that, in their mixtures with $n$-alkanes, interactions are mainly dispersive.

Systems including *n*-alkanoates with $v$ = 3, 4, or 5 or *n*-alkanoates with $u$ = constant also show decreasing positive $H_m^E$ values when the ester size increases (Tables 1, S2 and Figure 4), which can be explained in similar terms that above. For example, in the case of solutions with alkyl propanoates, $H_m^E$ ($u$ = 2)/J mol$^{-1}$ = 1424 ($v$ =1) [34]; 1127 ($v$ = 2) [45]; 893 ($v$ =3) [54]; 760 ($v$ = 4) [55]; 647 ($v$ = 5) [56] and $H_{m1}^{E,\infty}$ ($u$ = 2)/kJ mol$^{-1}$ = 5.4 ($v$ =1) [34]; 4.8 ($v$ =2) [45]; 4.4 ($v$ =3) [54]; 4.3 ($v$ = 4) [55]; 3.4 ($v$ = 5) [56].

*4.1.2   Excess molar volumes*

Values of $V_m^E$ for mixtures with methyl or ethyl *n*-alkanaote are represented in Figure 5 (see Table 2). Systems including a methyl *n*-alkanoate show $V_m^E$ results which are negative from $u \geq 7$. Since the corresponding values of $H_m^E$ are positive, this means that, for such solutions, the contribution to $V_m^E$ from structural effects is dominant over that due to interactional effects. A similar trend is observed for solutions with ethyl alkanoates. Slightly negative $V_m^E$ values are also encountered for the system containing pentyl pentanoate ($-0.050$ cm$^3$ mol$^{-1}$ [56]).

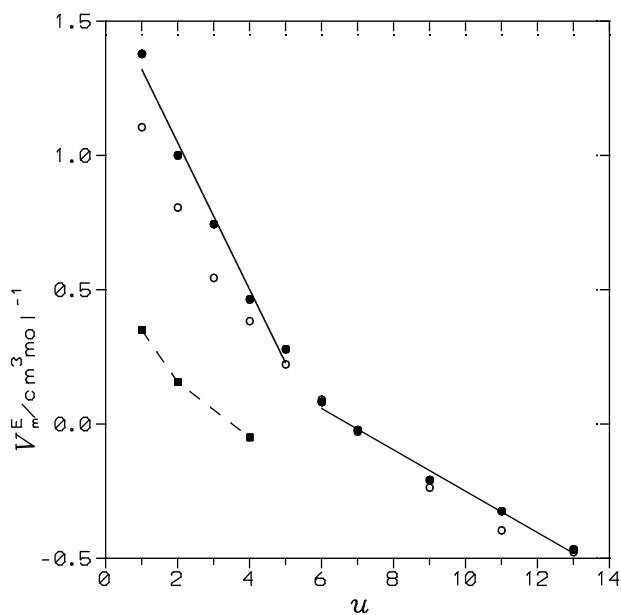

**Figure 5**. $V_m^E$ for CH$_3$(CH$_2$)$_{u-1}$COO(CH$_2$)$_{v-1}$CH$_3$ (1) + heptane (2) systems at equimolar composition and 298.15 K. Symbols: (●), $v$ =1; (O), $v$ =2; (■), $v$ =5; Each solid line is the result from the linear fitting of the data for mixtures with ($u$ = 1-5; $v$ =1) and with ($u$ = 6-13; $v$ =1), respectively. Dashed line is for the aid of the eye. For source of data, see Table 1 and reference [13].

It must be underlined that, along a given homologous series, e.g., mixtures with methyl $n$-alkanoates, $V_m^E$ also decreases when the ester size is increased (Figure 5). This can be explained taking into account: (i) the positive interactional contribution to $V_m^E$ decreases at this condition ($H_m^E$ varies similarly); (ii) a more negative contribution to $V_m^E$ from free volume effects, as it is indicated by the variation of the magnitude ($\alpha_p$(ester) $-\alpha_p$(heptane)) (Figure S2); (iii) the value of $|V_{m1} - V_{m2}|$ also increases from $u \geq 5$ and this also leads to a more negative contribution to $V_m^E$. In addition, the variation of this excess function is sharper for systems involving esters with $u \leq 5$ ($v$ = 1,2) and much smoother for solutions with $u \geq 6$ ($v$ =1,2) (Figure 5). If one takes into account that structural effects are dominant for systems with longer alkanoates, and that increase with $u$, the latter variation of $V_m^E$ can be explained in terms of an extra positive contribution to $V_m^E$ due to the breaking of stronger interactions between molecules of such $n$-alkanoates, which might be related to the existence of quasi-cyclic structures of the mentioned esters.

*4.2    Systems with a given n-alkanoate*

At this condition, both excess functions $H_m^E$ and $V_m^E$ increase with $n$ (Tables 1 and S2) which is, say, the normal behavior. The important point is that a number of mixtures show positive $H_m^E$ values and large and negative $V_m^E$ results, which reveals the existence of strong structural effects. This occurs for some mixtures with pentane. For example, $H_m^E$/J mol$^{-1}$ = 631 ($u$ = 6; $v$ =1) [57]; 449 ($u$ = 9, $v$ =1) [57], 305 ($u$ = 13,$v$ = 1) [35]; 302 ($u$ = 6, $v$ = 2) [45], 151 ($u$ = 9, $v$ = 2) [46]; 74 ($u$ =13, $v$ = 2) [46]; 638 ($u$ = 1; $v$ = 5); 323 ($u$ = 4; $v$ = 5) [56], and $V_m^E$/cm$^3$ mol$^{-1}$ = $-0.639$ ($u$ = 6; $v$ = 1) [58]; $-1.070$ ($u$ = 9, $v$ =1) [59], $-1.450$ ($u$ =13,$v$ =1) [59]; $-0.757$ ($u$ = 6, $v$ = 2) [45], $-1.200$ ($u$ = 9, $v$ = 2) [46]; $-1.472$ ($u$ =13, $v$ = 2) [46]; $-0.453$ ($u$ = 1; $v$ = 5) [56]; $-1.026$ ($u$ = 4; $v$ = 5) [56].

*4.3    Molar excess internal energy at constant volume.*

Since structural effects may become very important in the investigated mixtures, and such effects contribute meaningfully to $H_m^E$, it is pertinent to pay attention to the excess function $U_{Vm}^E$, which has been determined from the equation [60]:

$$U_{Vm}^E = H_m^E - T\frac{\alpha_p}{\kappa_T}V_m^E \qquad (10)$$

where $\alpha_p$ and $\kappa_T$ are, respectively, the isobaric thermal expansion coefficient and the coefficient of isothermal compressibility of the considered system. The $T\frac{\alpha_p}{\kappa_T}V_m^E$ term describes the contribution from the equation of state (eos) term to $H_m^E$. When the needed data are not available, $\alpha_p$ and $\kappa_T$ values can be calculated assuming that the mixtures behave ideally with regards to these properties. That is, $M^{id} = \phi_1 M_1 + \phi_2 M_2$; with $M_i = \alpha_{pi}$, or $\kappa_{T_i}$ and, $\phi_i = x_i V_{mi}/(x_1 V_{m1} + x_2 V_{m2})$. This approximation is commonly acceptable. Results are listed in Tables 1, S2, and S5 (see Figures 6,7, S3). From these results some statements can be provided. (i) In the systems methyl or ethyl alkanoate + heptane, $U_{Vm}^E > H_m^E$ from $u \geq 6$ ($v = 1$) or from $u \geq 7$ ($v = 2$). This clearly reveals the relevance of the eos contribution to $H_m^E$. (ii) $U_{Vm}^E$ (heptane) decreases when $u$ is increased ($v = 1,2$), but for larger $u$ values, this magnitude varies smoothly ($v =1$) or remains nearly constant ($v = 2$) (Table 1, Figure 4). (iii) The values of $U_{Vm1}^{E,\infty}$ (heptane) (determined from $U_{Vm}^E$ values over the whole concentration range) follow similar patterns that those of $H_{m1}^{E,\infty}$ and are more or less constant from $u \geq 7$. Thus, $U_{Vm1}^{E,\infty}$ (heptane)/kJ mol$^{-1}$ $\approx 3$ ($u \geq 7; v = 1$); 2.1 ($u \geq 7; v = 2$) (Table S3). This suggests that, in such mixtures, the increasing of the steric hindrance of the COO group is more relevant than interactional effects on $U_{Vm}^E$ values. Data for systems with alkyl ethanoate or propanoate or butanoate are scarce. Nevertheless, it seems clear that interactions between ester molecules become weaker when replacing, e.g., methyl by pentyl alkanoate. For example, $U_{Vm1}^{E,\infty}$ (heptane)/kJ mol$^{-1}$ = 5.7 (methyl ethanoate); 3.7 (pentyl ethanoate) (Table S3). (v) An important result arises from the variation of $U_{Vm}^E$ with $n$ in systems with a given ester (Tables S2, S5, Figures 6,7 S3). In such a case, $U_{Vm}^E$ increases with $n$ in systems with methyl butanoate or propyl propanoate ($u =1,2,3$), while the $U_{Vm}^E(n)$ function shows a minimum for systems with CH$_3$(CH$_2$)$_{u-1}$COO(CH$_2$)$_{v-1}$CH$_3$ from $u \geq 4$ ($v =1$) or from $u \geq 7$ ($v=2$), or from $u \geq 1$ ($v=4,5$). A similar dependence for $U_{Vm}^E(n)$ was encountered when investigating $n$-alkane mixtures with cyclic molecules such as cyclohexane, or benzene or cyclohexyl, or tetralin or cyclohexylbenzene [22]. That is, the existence of a minimum in the $U_{Vm}^E(n)$ function for a given homologous series may be due to the involved $n$-alkanoate forms quasi-cyclic structures, probably by folding of the hydrocarbon chain on >C=O. The present treatment extends results from Dusart et al. using infrared techniques [16,17] and shows that the formation of quasi-cyclic structures of esters in $n$-alkane medium is a rather common behavior. Our study allows conclude that this phenomenon occurs weakly in systems with propyl ethanoate [16,17], and that it might be present in solutions with methyl butanoate or ethyl heptanoate (Table S2 and S5).

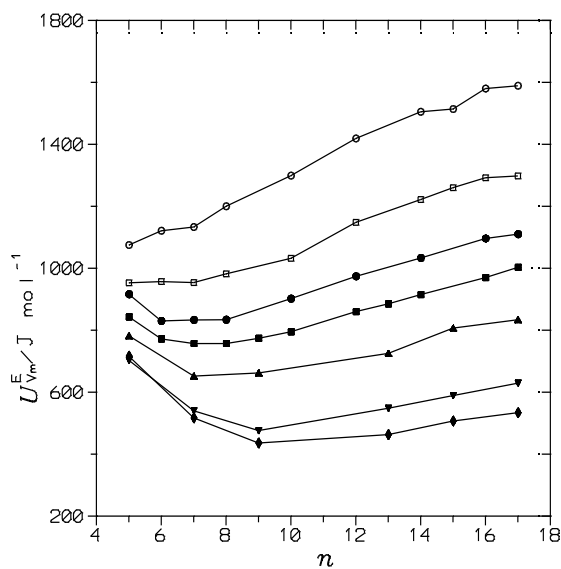

**Figure 6**. $U_{Vm}^E$ of $CH_3(CH_2)_{u-1}COOCH_3$ (1) + $n$-alkane (2) systems at equimolar composition and 298.15 K vs. $n$, the number of C atoms in the $n$-alkane. Symbols: (O), $u$ = 2; (□), $u$ = 3; (●), $u$ = 4; (■), $u$ = 5; (▲), $u$ = 7; (▼), $u$ = 11; (♦), $u$ = 13. Lines are for the aid of the eye. Numerical data are listed in Tables S2 and S5.

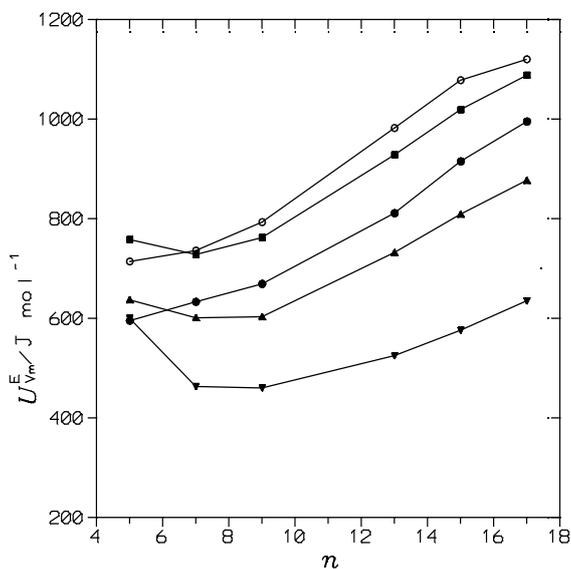

**Figure 7**. $U_{Vm}^E$ of $CH_3(CH_2)_{u-1}COO(CH_2)_{v-1}CH_3$ (1) + $n$-alkane (2) systems at equimolar composition and 298.15 K vs. $n$, the number of C atoms in the $n$-alkane. Symbols: (O), $u$ = 1; $v$ =3; (●), $u$ = 3; $v$ =3; (■), $u$ = 1; $v$ =5; (▲), $u$ = 2; $v$ =5; (▼), $u$ = 4; $v$ =5. Lines are for the aid of the eye. Numerical data are listed in Tables S2 and S5.

*4.4    Dynamic viscosities*

Deviations of dynamic viscosity ($\Delta \eta = \eta - (x_1\eta_1 + x_2\eta_2)$) of $n$-alkanoate + $n$-alkane mixtures are negative (Table 2, Figure 8), which is consistent with the fact that no specific interactions exist between the system compounds [61,62]. The $\Delta\eta$ values of $CH_3(CH_2)_{u-1}COOCH_3$ + heptane mixtures decrease when $u$ is increased (Table 3)). The same trend is observed for $H_m^E$ (or $U_{Vm}^E$) and $V_m^E$. It has been suggested that a certain correlation exists between $\Delta\eta$ and $V_m^E$, these functions having opposite signs [63,64]. Indeed, this is held for systems with $u$ = 1,3,5. However, the system containing methyl n-decanoate has negative values of $\Delta\eta$ (−0.247 mPa [65]) and of $V_m^E$ (−0.154 cm$^3$mol$^{-1}$ [59]). It suggests that there is no clear relationship between such thermophysical magnitudes. On the other hand, the decreasing of $U_{Vm}^E$ when $u$ increases has been above related with a lower number of broken interactions between like molecules, and it cannot explain the corresponding increased negative values of $\Delta\eta$.

The Bloomfield-Dewans's model [24], assuming that $\beta$ = 0, provides excellent results (Table 2) if one considers that they are obtained from volumetric data. Unfortunately, from this analysis, nothing can be stated about the possible formation of quasi-cyclic structures of n-alkanoates. In order to continue investigating this point, now it is paid attention to ester (fixed) + n-alkane systems. The data are scarce since they are only available for mixtures including methyl ethanoate, or butyl ethanoate or propyl propanoate or methyl decanaote (see Figure 8).

Inspection of Figure 8 shows that $\Delta\eta$ is nearly constant for lower $n$ values in the case of solutions with methyl acetate, decreasing for larger $n$ values. The dependence of $\Delta\eta$ with $n$ is somewhat different for the mixtures including butyl ethanoate or methyl decanoate. In fact, $\Delta\eta$ increases with $n$ and decreases for larger $n$ values. The large negative $\Delta\eta$ result for the methyl ethanoate + hexadecane mixture can be explained taking into account that, at 298.15 K, this system is not far from its UCST which means that a large number of interactions between like molecules are broken upon mixing. It is remarkable that, in the case of systems with butyl ethanoate or methyl decanoate, the variation of $\Delta\eta$ ($n$) is consistent with that of $U_{Vm}^E$ ($n$). This excess function decreases for low $n$ values indicating that a lower number of ester-ester interactions are disrupted (consequently, $\Delta\eta$ increases) and increases for large $n$ values, which might be due to the quasi-cyclic structures of the alkanoate are good breakers of the correlations of molecular orientations of longer n-alkanes, leading to increased negative values of $\Delta\eta$. However, this statement should be taken with caution until more viscosity data for homologous series are available.

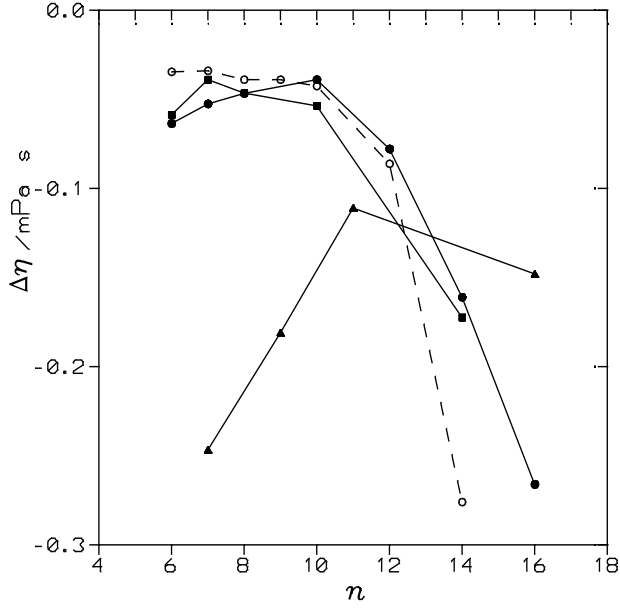

**Figure 8**. $\Delta\eta$ results for $CH_3(CH_2)_{u-1}COO(CH_2)_{v-1}CH_3$ (1) + heptane (2) systems at 298.15 K. Symbols, experimental results (for source of data, see Table 2 and [82]): (O), ($u$ = 1, $v$ = 1); (●), ($u$ = 1, $v$ = 4); (■), ($u$ = 2, $v$ = 3); (▲); ($u$ = 9, $v$ =1). Lines are for the aid of the eye.

With regards to the results obtained from the application of the Grunberg-Nissan equation [23], we note that it correlates very well the viscosity data under consideration (Table 2). The $G_{12}$ parameter is usually negative, indicating that dispersive interactions are dominant [62,66,67], in agreement with our previous findings. Positive values of the parameter are encountered for the systems: methyl decanoate + heptane, or methyl ethanoate + hexadecane, or propyl propanoate + tetradecane (Table 2), which may be ascribed to size effects. In order to investigate this point, we have determined deviations of molar Gibbs energy of activation from the linear dependence with composition, $\Delta(\Delta G_m^*)$, according to the equation [68]:

$$\Delta(\Delta G_m^*) = \Delta G_m^* - x_1 \Delta G_{m1}^* - x_2 \Delta G_{m2}^* = RT[\ln(V_m\eta) - x_1\ln(V_{m1}\eta_1) - x_2\ln(V_{m2}\eta_2)] \qquad (11)$$

Positive results of this magnitude have been also obtained for the same solutions mentioned above. Mixtures of benzene [68] or cyclohexane [69] and longer $n$-alkanes, or formed by long chain 1-alkanol and an amine [70,71] are also characterized by positive $\Delta(\Delta G_m^*)$ results, which underlines the importance of size effects on this magnitude. Finally, it must be mentioned that for the series $CH_3(CH_2)_{u-1}COOCH_3$ + heptane, the function $\Delta(\Delta G_m^*)$ shows a minimum at $u$ =3, in agreement with the finding of Dusart et al., who demonstrated the existence of quasi-cycles of methyl butanaote from viscosity data [20].

*4.5   Flory model*

Inspection of Tables 1 and S2 shows that, for many of the considered systems, deviations between experimental and theoretical values of $H_m^E$ are ≤ 10%, which suggests the existence of weak orientational effects. Large values of $\sigma_r(H_m^E)$ are obtained for systems with pentane or for the mixtures ethyl tetradecanoate + heptane, or + nonane. For example, $\sigma_r(H_m^E)$(pentane) = 0.189 (methyl hexanoate); 0.269 (methyl decanoate); 0.354 (methyl tetradecanoate). Since orientational effects become weaker when the ester size increases along a given homologous series, the poorer results obtained for systems with methyl or ethyl decanoate, or methyl or ethyl tetradecanoate (Tables 1 and S2) cannot be ascribed to such effects, but to the existence of strong structural effects which largely contribute $H_m^E$ (eos term), and that are not properly described by the model. This can be shown comparing $H_m^E$ and $U_{Vm}^E$ results (Tables 1, S2, S5). Thus, $H_m^E$ (pentane)/J mol$^{-1}$ = 305 (methyl tetradecanoate) [35], 74 (ethyl tetradecanote) [46] and, in the same sequence, $U_{Vm}^E$ (pentane)/J mol$^{-1}$= 709; 482. In order to examine the influence of the mentioned structural effects on Flory results, the model was applied to mixtures with pentane, heptane, or pentadecane using the corresponding $U_{Vm}^E$ data. Results are listed in Table 1 (see Figure 2). A meaningful improvement is obtained for systems with pentane. We have determined the average standard relative deviation for $H_m^E$ and $U_{Vm}^E$, $\bar{\sigma}_r(F_m^E)$ ($F_m^E = H_m^E, U_{Vm}^E$), for each homologous series, according to the expression:

$$\bar{\sigma}_r(F_m^E) = \frac{1}{N_S}\sum \sigma_r(F_m^E) \qquad (12)$$

where $N_S$ is the number of systems of each series. The results are: $\bar{\sigma}_r(H_m^E)$(pentane) = 0.257 (methyl alkanoate); 0.330 (ethyl alkanoate); 0.287 (propyl alkanoate); 0.223 (butyl alkanoate); 0.164 (pentyl alkanoate), and in the same order, $\bar{\sigma}_r(U_{Vm}^E)$ (pentane) = 0.133; 0.104; 0.228; 0.133, 0.073. For mixtures containing heptane or pentadecane, results are similar. It must be mentioned that the application of group contribution models for the characterization of the COO/aliphatic contacts in mixtures where strong structural effects may be present is quite difficult since the $H_m^E$ values can largely change when increasing the alkane size in systems with a given ester. Compare, e.g, $H_m^E$ (ethyl tetradecanoate)/J mol$^{-1}$ = 74 (pentane); 231 (heptane), and 414 (heptadecane) [46]. Thus, solutions with pentane were discarded in an application of the UNIFAC model to mixtures with butyl alkanoate [55]. In this framework, it should be mentioned that the Dortmund version of UNIFAC and DISQUAC models fail when describing $H_\text{m}^\text{E}$ values of mixtures such as pentyl pentanoate + pentane [56].

With regards to Flory results on $V_\text{m}^\text{E}$, the theory usually predicts the increase of this magnitude with $n$ in systems involving a fixed $n$-alkanoate. However, for a number of systems, a decrease of the $V_\text{m}^\text{E}$ is observed when replacing pentadecane by heptadecane (see results for mixtures with ethyl heptanoate, or pentyl pentanoate, Table S2). That is, the model overestimates the structural contribution to $V_\text{m}^\text{E}$ in systems with heptadecane.

## 5. Conclusions

Interactions in the studied mixtures are of dispersive type. Large structural effects are encountered in a number of systems, e.g, in those containing pentane. The variations of $\Delta\Delta H_\text{vap,i}$ and of $D_\text{ce,i}$ with the ester size along homologous series formed by methyl or ethyl $n$-alkanoates reveal the existence of some structural change in longer $n$-alkanoates which leads to stronger interactions between them. This is supported by the variation of $V_\text{m}^\text{E}$ values of the corresponding heptane mixtures. The observed decrease of $H_\text{m}^\text{E}$ for this type of systems seems to be more related to the COO group is more sterically hindered than to interactional effects. The $U_{V\text{m}}^\text{E}(n)$ function shows a minimum for systems with esters characterized by ($u \geq 4$, $v = 1$); ($u \geq 7$, $v=2$), or ($u \geq 1$, $v=5$). That is, certain $n$-alkanoates, in an alkane medium, can form quasi-cyclic structures. Free volume effects are determinant for describing viscosity data of these mixtures. It is remarkable that, for systems with butyl ethanoate or methyl decanoate, the variation of $\Delta\eta$ is consistent with that of $U_{V\text{m}}^\text{E}$, which suggests that the quasi-cycles of the alkanoate are good breakers of the correlations of molecular orientations of longer $n$-alkanes. Results from the Flory model are rather poor for those systems with large structural effects. Better results for these mixtures are obtained when the model is applied using $U_{V\text{m}}^\text{E}$ data.

**Acknowledgements**

This work was carried out under "Project PID2022-137104NA-I00", funded by MICIN/AEI/10.13039/501100011033/ and by FEDER, UE.

SUPPLEMENTARY INFORMATION

# *N*-ALKANOATE + *N*-ALKANE MIXTURES: FOLDING OF HYDROCARBON CHAINS OF *N*-ALKANOATES

Juan Antonio González,* Fernando Hevia, Luis Felipe Sanz, Daniel Lozano-Martín, Isaías. García de la Fuente . José Carlos Cobos

ªG.E.T.E.F., Departamento de Física Aplicada, Facultad de Ciencias, Universidad de Valladolid, Paseo de Belén, 7, 47011 Valladolid, Spain.

*Corresponding author, e-mail: jagl@termo.uva.es

**TABLE 1**

Physical constants and Flory parameters[a] of pure compounds at $T = 298.15$ K.

| Compound | $V_{\text{mi}}$/cm$^3$ mol$^{-1}$ | $\alpha_{pi}$/$10^{-3}$ K$^{-1}$ | $\kappa_{Ti}$/$10^{-12}$ Pa$^{-1}$ | $V_{\text{mi}}^*$/cm$^3$ mol$^{-1}$ | $P_i^*$/ J cm$^{-3}$ |
|---|---|---|---|---|---|
| methyl hexanoate[b] | 148.05 | 1.08 | 944 | 117.14 | 544.9 |
| methyl heptanoate | 164.73[c] | 1.04[c] | 889[d] | 131.17 | 550 |
| methyl octanoate[b] | 181.41 | 0.997 | 869 | 145.47 | 532 |
| methyl decanoate[b] | 214.56 | 0.939 | 814 | 173.72 | 524.6 |
| methyl dodecanoate[b] | 247.81 | 0.895 | 755 | 202.17 | 531 |
| methyl tetradecanoate[b] | 280.91 | 0.867 | 709 | 230.29 | 542 |
| ethyl pentanoate[e] | 149.52 | 1.08 | 992 | 118.30 | 518.5 |
| ethyl hexanoate[f] | 166.48 | 1.10 | 981 | 131.31 | 537.4 |
| ethyl heptanoate[f] | 183.09 | 1.04 | 934 | 145.79 | 523.6 |
| ethyl octanoate[g] | 199.80 | 1.02 | 904 | 159.61 | 527 |
| ethyl decanoate[g] | 232.57 | 0.96 | 848 | 187.96 | 518.5 |
| ethyl dodecanoate[g] | 266.11 | 0.94 | 814 | 215.43 | 525 |
| ethyl tetradecanoate[g] | 282.81 | 0.91 | 782 | 230.12 | 524 |
| propyl butanoate | 149.99[h] | 1.15[h,i] | 1006[d] | 117.39 | 556.4 |
| butyl butanoate | 166.78[j] | 1.04[j,k] | 911[d] | 132.80 | 536.8 |
| pentyl propanoate | 166.18[l] | 1.08[l,m] | 939[d] | 131.48 | 547.8 |
| pentyl pentanoate | 200.16[l] | 1.02[l,k] | 892[d] | 159.9 | 534.2 |
| tridecane[n] | 244.91 | 0.95 | 944 | 197.93 | 459.1 |
| pentadecane[n] | 277.66 | 0.91 | 882 | 225.93 | 464.5 |
| heptadecane[o] | 310.96 | 0.899 | 819 | 253.51 | 492.4 |

[a]$V_{\text{mi}}$, molar volume; $\alpha_{pi}$, isobaric expansion coefficient; $\kappa_{Ti}$, isothermal compressibility, $V_{\text{mi}}^*$, $P_i^*$, reduction parameters for volume and pressure, respectively; [b][S1]; [c]S2]; [d]estimated using the Manzini-Creszenci method [S3]; [e][S4]; [f][S5]; [g][S6]; [h][S7]: [i][S8]; [j][S9]; [k][S10]; [l][S9]; [m][S11]; [n][S12]; [o][S13].

TABLE S2

Excess molar enthalpies, $H_m^E$, volumes, $V_m^E$, and isochoric internal energies, $U_{Vm}^E$, at equimolar composition and 298.15 K, for n-alkanoate (1) + n-alkane (2) mixtures. Values of the Flory interaction parameter, $X_{12}$, and of the relative standard deviations for $H_m^E$, $\sigma_r(H_m^E)$, (equation 7), are also included.

| $n$-$C_n$ | $N^a$ | $H_m^E$ / J mol$^{-1}$ | $X_{12}$ / J cm$^{-3}$ | $\sigma_r(H_m^E)$ | Ref. | $V_m^E$ / cm$^3$ mol$^{-1}$ | | Ref. | $U_{Vm}^E$ / J mol$^{-1}$ |
|---|---|---|---|---|---|---|---|---|---|
| | | | | | | Exp | Flory | | |
| methyl heptanoate + n-alkane | | | | | | | | | |
| 5 | 21 | 631 | 23.00 | 0.193 | S14 | −0.639 | −0.540 | S15 | 800 |
| 6 | 20 | 684 | 22.58 | 0.080 | S16 | −0.139 | −0.117 | S16 | 722 |
| 7 | 18 | 741 | 22.94 | 0.073 | S17 | 0.082 | 0.244 | S15 | 718 |
| 8 | 17 | 778 | 22.97 | 0.081 | S18 | 0.309 | 0.424 | S18 | 685 |
| 9 | 17 | 845 | 24.00 | 0.074 | S19 | 0.405 | 0.593 | S15 | 721 |
| 10 | 20 | 886 | 24.39 | 0.092 | S20 | 0.487 | 0.666 | S20 | 735 |
| 12 | 20 | 967 | 25.51 | 0.077 | S21 | 0.605 | 0.838 | S21 | 778 |
| 13 | 18 | 1008 | 26.08 | 0.069 | S22 | 0.665 | 0.847 | S15 | 796 |
| 14 | 21 | 1050 | 26.72 | 0.084 | S23 | 0.671 | 0.933 | S23 | 835 |
| 15 | 18 | 1101 | 27.59 | 0.062 | S24 | 0.701 | 0.924 | S25 | 874 |
| 16 | 18 | 1115 | 27.59 | 0.067 | S26 | 0.732 | 0.973 | S26 | 880 |
| 17 | 18 | 1147 | 27.93 | 0.085 | S27 | 0.739 | 0.859 | S27 | 899 |
| methyl octanoate + n-alkane | | | | | | | | | |
| 5 | 19 | 561 | 19.76 | 0.189 | S14 | −0.826 | −0.689 | S1 | 778 |
| 7 | 17 | 645 | 18.93 | 0.089 | S17 | −0.023 | 0.090 | S1 | 652 |
| 9 | 16 | 756 | 20.14 | 0.085 | S19 | 0.309 | 0.450 | S1 | 662 |
| 13 | 18 | 902 | 21.67 | 0.055 | S22 | 0.558 | 0.702 | S1 | 725 |
| 15 | 16 | 1001 | 23.26 | 0.064 | S24 | 0.608 | 0.787 | S25 | 806 |
| 17 | 18 | 1046 | 23.55 | 0.057 | S27 | 0.638 | 0.743 | S27 | 834 |
| methyl decanoate + n-alkane | | | | | | | | | |
| 5 | 19 | 449 | 15.14 | 0.269 | S14 | −1.070 | −1.005 | S1 | 739 |
| 7 | 20 | 550 | 14.81 | 0.106 | S17 | −0.209 | −0.154 | S1 | 612 |
| 9 | 16 | 639 | 15.32 | 0.125 | S19 | 0.144 | 0.225 | S1 | 595 |
| 12 | 12 | 722 | 15.72 | 0.146 | S28 | 0.366 | 0.476 | S1 | 607 |
| 13 | 18 | 756 | 16.08 | 0.092 | S22 | 0.404 | 0.510 | S1 | 627 |
| 14 | 12 | 789 | 16.46 | 0.085 | S28 | 0.440 | 0.600 | S28 | 648 |

TABLE S2 (continued)

| | | | | | | | | | |
|---|---|---|---|---|---|---|---|---|---|
| 15 | 18 | 831 | 17.00 | 0.061 | S24 | 0.460 | 0.596 | S25 | 683 |
| 16 | 12 | 829 | 16.68 | 0.071 | S28 | 0.492 | 0.629 | S28 | 671 |
| 17 | 18 | 885 | 17.51 | 0.070 | S27 | 0.512 | 0.597 | S27 | 714 |
| methyl dodecanoate + $n$-alkane | | | | | | | | | |
| 5 | 17 | 357 | 11.92 | 0.348 | S29 | −1.251 | −1.139 | S1 | 700 |
| 7 | 17 | 442 | 11.26 | 0.096 | S29 | −0.325 | −0.436 | S1 | 537 |
| 9 | 20 | 515 | 11.40 | 0.110 | S30 | 0.114 | −0.243 | S1 | 476 |
| 13 | 18 | 640 | 12.35 | 0.104 | S22 | 0.283 | 0.332 | S1 | 548 |
| 15 | 16 | 707 | 13.04 | 0.077 | S24 | 0.354 | 0.435 | S25 | 589 |
| 17 | 18 | 749 | 13.31 | 0.086 | S27 | 0.393 | 0.465 | S27 | 629 |
| methyl tetradecanoate + $n$-alkane | | | | | | | | | |
| 5 | 18 | 305 | 10.11 | 0.354 | S29 | −1.450 | −1.609 | S1 | 709 |
| 7 | 18 | 373 | 9.12 | 0.118 | S29 | −0.467 | −0.670 | S1 | 516 |
| 9 | 20 | 425 | 8.85 | 0.128 | S30 | −0.097 | −0.233 | S1 | 456 |
| 13 | 18 | 531 | 9.46 | 0.092 | S22 | 0.191 | 0.162 | S1 | 463 |
| 15 | 18 | 597 | 10.12 | 0.085 | S24 | 0.273 | 0.286 | S25 | 506 |
| 17 | 20 | 639 | 10.42 | 0.094 | S27 | 0.307 | 0.344 | S27 | 534 |
| ethyl heptanoate + $n$-alkane | | | | | | | | | |
| 5 | 17 | 302 | 11.21 | 0.232 | S5 | −0.757 | −0.746 | S5 | 506 |
| 7 | 17 | 530 | 15.43 | 0.070 | S5 | 0.091 | 0.122 | S5 | 510 |
| 9 | 16 | 650 | 17.23 | 0.089 | S5 | 0.325 | 0.460 | S5 | 552 |
| 13 | 17 | 834 | 20.04 | 0.086 | S5 | 0.578 | 0.698 | S5 | 653 |
| 15 | 17 | 944 | 21.95 | 0.095 | S5 | 0.645 | 0.774 | S5 | 739 |
| 17 | 16 | 1022 | 23.06 | 0.062 | S5 | 0.701 | 0.726 | S5 | 791 |
| ethyl octanoate + $n$-alkane | | | | | | | | | |
| 5 | 18 | 234 | 8.77 | 0.258 | S6 | −0.958 | −0.914 | S6 | 491 |
| 7 | 17 | 437 | 12.16 | 0.041 | S6 | −0.029 | −0.028 | S6 | 445 |
| 9 | 16 | 535 | 13.39 | 0.083 | S6 | 0.253 | 0.322 | S6 | 458 |
| 13 | 17 | 695 | 15.66 | 0.102 | S6 | 0.497 | 0.579 | S6 | 538 |
| 15 | 17 | 783 | 17.07 | 0.129 | S6 | 0.572 | 0.654 | S6 | 599 |
| 17 | 17 | 876 | 18.50 | 0.202 | S6 | 0.619 | 0.631 | S6 | 671 |
| ethyl decanoate + $n$-alkane | | | | | | | | | |
| 5 | 17 | 151 | 6.19 | 0.316 | S6 | −1.200 | −1.174 | S6 | 482 |
| 7 | 19 | 300 | 7.92 | 0.035 | S6 | −0.237 | −0.284 | S6 | 370 |
| 9 | 19 | 356 | 8.14 | 0.075 | S6 | 0.079 | 0.065 | S6 | 332 |

TABLE S2 (continued)

| | | | | | | | | | |
|---|---|---|---|---|---|---|---|---|---|
| 13 | 17 | 499 | 10.06 | 0.140 | S6 | 0.343 | 0.362 | S6 | 390 |
| 15 | 17 | 582 | 11.29 | 0.106 | S6 | 0.435 | 0.449 | S6 | 444 |
| 17 | 17 | 650 | 12.19 | 0.086 | S6 | 0.491 | 0.451 | S6 | 487 |
| ethyl dodecanoate + *n*-alkane | | | | | | | | | |
| 5 | 13 | 99 | 4.65 | 0.336 | S6 | −1.332 | −1.330 | S6 | 467 |
| 7 | 18 | 251 | 6.31 | 0.115 | S6 | −0.403 | −0.430 | S6 | 375 |
| 9 | 16 | 297 | 6.31 | 0.110 | S6 | −0.048 | −0.054 | S6 | 312 |
| 13 | 17 | 385 | 7.09 | 0.111 | S6 | 0.232 | 0.248 | S6 | 310 |
| 15 | 18 | 437 | 7.72 | 0.132 | S6 | 0.314 | 0.331 | S6 | 333 |
| 17 | 17 | 476 | 8.12 | 0.130 | S6 | 0.379 | 0.333 | S6 | 349 |
| ethyl tetradecanoate + *n*-alkane | | | | | | | | | |
| 5 | 10 | 74 | 4.09 | 0.639 | S6 | −1.472 | −1.500 | S6 | 482 |
| 7 | 17 | 231 | 5.79 | 0.224 | S6 | −0.477 | −0.540 | S6 | 375 |
| 9 | 17 | 276 | 5.73 | 0.169 | S6 | −0.146 | −0.149 | S6 | 231 |
| 13 | 17 | 348 | 6.16 | 0.106 | S6 | 0.176 | 0.169 | S6 | 291 |
| 15 | 18 | 377 | 6.37 | 0.126 | S6 | 0.245 | 0.246 | S6 | 297 |
| 17 | 18 | 414 | 6.74 | 0.102 | S6 | 0.296 | 0.271 | S6 | 314 |
| propyl propanoate + *n*-alkane | | | | | | | | | |
| 5 | 17 | 670 | 26.96 | 0.249 | S7 | −0.168 | −0.263 | S7 | 714 |
| 7 | 17 | 893 | 31.76 | 0.091 | S7 | 0.562 | 0.668 | S7 | 736 |
| 9 | 17 | 1019 | 34.15 | 0.075 | S7 | 0.753 | 1.089 | S7 | 793 |
| 13 | 16 | 1269 | 39.55 | 0.064 | S7 | 0.932 | 1.448 | S7 | 982 |
| 15 | 17 | 1390 | 42.21 | 0.070 | S7 | 0.981 | 1.564 | S7 | 1078 |
| 17 | 17 | 1464 | 43.38 | 0.068 | S7 | 1.041 | 1.457 | S7 | 1120 |
| propyl butanoate + *n*-alkane | | | | | | | | | |
| 5 | 19 | 494 | 18.77 | 0.354 | S7 | −0.379 | −0.336 | S7 | 595 |
| 7 | 16 | 747 | 24.53 | 0.091 | S7 | 0.393 | 0.508 | S7 | 633 |
| 9 | 17 | 855 | 26.07 | 0.080 | S7 | 0.616 | 0.805 | S7 | 669 |
| 13 | 17 | 1065 | 30.07 | 0.064 | S7 | 0.803 | 1.019 | S7 | 811 |
| 15 | 17 | 1192 | 32.76 | 0.046 | S7 | 0.865 | 1.097 | S7 | 915 |
| 17 | 17 | 1294 | 34.53 | 0.059 | S7 | 0.901 | 1.011 | S7 | 995 |
| butyl propanoate + *n*-alkane | | | | | | | | | |
| 5 | 15 | 577 | 21.91 | 0.193 | S31 | −0.484 | −0.290 | S31 | 706 |
| 7 | 18 | 760 | 24.95 | 0.100 | S31 | 0.359 | 0.492 | S31 | 655 |
| 9 | 16 | 874 | 26.75 | 0.073 | S31 | 0.571 | 0.805 | S31 | 701 |

TABLE S2 (continued)

| | | | | | | | | | |
|---|---|---|---|---|---|---|---|---|---|
| 13 | 18 | 1094 | 30.87 | 0.080 | S31 | 0.791 | 1.032 | S31 | 843 |
| 15 | 18 | 1208 | 33.13 | 0.082 | S31 | 0.861 | 1.104 | S31 | 932 |
| 17 | 17 | 1318 | 35.18 | 0.074 | S31 | 0.922 | 1.033 | S31 | 1011 |
| butyl butanoate + *n*-alkane | | | | | | | | | |
| 5 | 15 | 400 | 15.10 | 0.320 | S31 | −0.632 | −0.690 | S31 | 568 |
| 7 | 17 | 565 | 17.43 | 0.159 | S31 | 0.190 | 0.131 | S31 | 510 |
| 9 | 17 | 679 | 19.15 | 0.084 | S31 | 0.407 | 0.475 | S31 | 556 |
| 13 | 18 | 918 | 23.55 | 0.057 | S31 | 0.679 | 0.769 | S31 | 703 |
| 15 | 17 | 1038 | 25.79 | 0.098 | S31 | 0.761 | 0.856 | S31 | 794 |
| 17 | 18 | 1160 | 27.98 | 0.087 | S31 | 0.804 | 0.844 | S31 | 893 |
| pentyl ethanoate + *n*-alkane | | | | | | | | | |
| 5 | 16 | 636 | 24.21 | 0.217 | S9 | −0.453 | −0.211 | S9 | 758 |
| 7 | 17 | 828 | 27.19 | 0.086 | S9 | 0.350 | 0.532 | S9 | 728 |
| 9 | 18 | 947 | 28.84 | 0.097 | S9 | 0.612 | 0.807 | S9 | 761 |
| 13 | 16 | 1186 | 33.13 | 0.073 | S9 | 0.814 | 1.010 | S9 | 928 |
| 15 | 20 | 1305 | 35.43 | 0.081 | S9 | 0.882 | 1.068 | S9 | 1019 |
| 17 | 15 | 1412 | 37.27 | 0.119 | S9 | 0.971 | 1.003 | S9 | 1088 |
| pentyl propanoate + *n*-alkane | | | | | | | | | |
| 5 | 16 | 469 | 17.19 | 0.104 | S9 | −0.631 | −0.559 | S9 | 635 |
| 7 | 17 | 647 | 19.89 | 0.078 | S9 | 0.156 | 0.265 | S9 | 601 |
| 9 | 17 | 739 | 20.76 | 0.087 | S9 | 0.449 | 0.584 | S9 | 603 |
| 13 | 18 | 945 | 24.47 | 0.061 | S9 | 0.670 | 0.849 | S9 | 732 |
| 15 | 16 | 1056 | 26.51 | 0.070 | S9 | 0.764 | 0.930 | S9 | 809 |
| 17 | 18 | 1154 | 28.18 | 0.061 | S9 | 0.830 | 0.879 | S9 | 877 |
| pentyl pentanoate + *n*-alkane | | | | | | | | | |
| 5 | 15 | 323 | 11.43 | 0.170 | S9 | −1.026 | −0.864 | S9 | 601 |
| 7 | 15 | 448 | 12.44 | 0.114 | S9 | −0.050 | −0.037 | S9 | 463 |
| 9 | 18 | 536 | 13.78 | 0.099 | S9 | 0.249 | 0.327 | S9 | 460 |
| 13 | 16 | 692 | 15.57 | 0.076 | S9 | 0.528 | 0.581 | S9 | 525 |
| 15 | 19 | 771 | 16.79 | 0.087 | S9 | 0.612 | 0.654 | S9 | 575 |
| 17 | 16 | 862 | 18.17 | 0.082 | S9 | 0.682 | 0.632 | S9 | 635 |

[a]number of data points

**TABLE S3**

Excess partial molar functions at infinite dilution of the first compound, $F_{m1}^{E,\infty}$, and at 298.15 K for *n*-alkanoate (1) + heptane (2) mixtures: $F = H$ (enthalpy) or $U_V$ (isochoric internal energy)

| *n*-alkanoate | $H_{m1}^{E,\infty}$ /kJ mol$^{-1}$ | Ref. | $U_{Vm1}^{E,\infty}$ /kJ mol$^{-1}$ |
|---|---|---|---|
| methyl ethanoate | 8.3 | S32 | 5.7 |
| methyl propanoate | 5.4 | S17 | 4.3 |
| methyl butanoate | 5.0 | S17 | 4.0 |
| methyl pentanoate | 4.5 | S17 | 3.8 |
| methyl hexanoate | 4.2 | S17 | 3.5 |
| methyl heptanoate | 4.0 | S17 | 3.4 |
| methyl octanoate | 3.5 | S17 | 3.1 |
| methyl decanoate | 3.3 | S17 | 3.1 |
| methyl dodecanoate | 2.7 | S29 | 3.0 |
| methyl tetradecanoate | 2.7 | S29 | 3.0 |
| ethyl ethanoate | 7.0 | S5 | 5.1 |
| ethyl propanoate | 4.8 | S5 | 3.7 |
| ethyl butanoate | 4.5 | S5 | 3.7 |
| ethyl pentanoate | 3.9 | S5 | 3.1 |
| ethyl hexanoate | 3.6 | S16 | 3.2 |
| ethyl heptanoate | 2.7 | S5 | 2.4 |
| ethyl octanoate | 2.0 | S6 | 2.1 |
| ethyl decanoate | 1.5 | S6 | 2.0 |
| ethyl dodecanoate | 1.5 | S6 | 2.0 |
| ethyl tetradecanoate | 1.8 | S6 | 2.5 |
| propyl ethanoate | 5.6 | S33 | 4.6 |
| propyl propanoate | 4.4 | S7 | 3.7 |
| propyl butanoate | 3.7 | S7 | 3.1 |
| butyl ethanoate | 5.4 | S31 | 4.5 |
| butyl propanoate | 4.3 | S31 | 3.8 |
| butyl butanoate | 3.6 | S31 | 3.4 |
| pentyl ethanoate | 4.5 | S9 | 3.7 |
| pentyl propanote | 3.4 | S7 | 3.2 |
| pentyl pentanoate | 2.9 | S9 | 3.0 |

# TABLE S4

Standard enthalpies of vaporization at 298.15 K, $\Delta H_{vap,i}$, $\Delta\Delta H_{vap,i}$ (= $\Delta H_{vap,i}$ (pure compound) − $\Delta H_{vap,i}$ (isomeric alkane)); molar volumes, $V_{mi}$, density of cohesive energy, $D_{cei}$ (equation 10), internal pressures, $P_{int,i}$ (equation 11)[a] and $q_i$ (= $P_{int,i} / D_{cei}$).

| compound | $\Delta H_{vap,i}$ / kJ mol$^{-1}$ | Ref. | $\Delta\Delta H_{vap,i}$ / kJ mol$^{-1}$ | $V_{mi}$ / cm$^3$ mol$^{-1}$ | Ref. | $D_{cei}$ / MPa | $P_{int,i}$ / MPa | $q_i$ |
|---|---|---|---|---|---|---|---|---|
| methyl ethanoate | 32.29 | S34 | | 79.82 | S1 | 373.5 | 370.3 | 0.99 |
| methyl butanoate | 39.28 | S34 | 12.8 | 114.42 | S1 | 321.6 | 352.7 | 1.10 |
| methyl pentanoate | 43.10 | S34 | 11.5 | 131.30 | S2 | 309.4 | 350.7 | 1.13 |
| methyl hexanoate | 48.04 | S34 | 11.5 | 148.05 | S1 | 307.7 | 341.1 | 1.11 |
| methyl heptanoate | 51.62 | S34 | 10.1 | 164.73 | S10 | 298.3 | 348.7 | 1.17 |
| methyl octanoate | 56.41 | S34 | 10.0 | 181.41 | S1 | 297.3 | 342.0 | 1.15 |
| methyl decanoate | 66.75 | S34 | 10.3 | 214.56 | S1 | 299.6 | 343.8 | 1.15 |
| methyl dodecanoate | 77.17 | S34 | 10.7 | 247.81 | S1 | 301.4 | 353.3 | 1.17 |
| methyl tetradecanoate | 86.98 | S34 | 10.9 | 280.90 | S1 | 300.8 | 364.6 | 1.21 |
| ethyl butanoate | 42.68 | S34 | 11.1 | 133.01 | S35 | 302.2 | 348.7 | 1.15 |
| ethyl pentanoate | 47.01 | S34 | 10.4 | 149.39 | S35 | 298.1 | 324.5 | 1.09 |
| ethyl hexanoate | 51.72 | S34 | 10.2 | 166.48 | S5 | 295.8 | 334.2 | 1.13 |
| ethyl heptanoate | 55.8 | S35 | 9.4 | 182.09 | S5 | 292.8 | 331.9 | 1.13 |
| ethyl octanoate | 59.5 | S36 | 8.1 | 199.80 | S6 | 285.4 | 336.3 | 1.18 |
| ethyl decanoate | 70.5 | S37 | 9.0 | 232.97 | S6 | 292.0 | 337.4 | 1.16 |
| ethyl dodecanoate | 80. | S38 | 8.7 | 266.11 | S6 | 291.3 | 344.2 | 1.18 |
| 2-butanone | 34.79 | S34 | | 90.14 | S39 | 385.8 | 333.7 | 0.86 |
| 2-pentanone | 38.46 | S34 | 12.0 | 107.46 | S40 | 334.8 | 324.9 | 0.97 |
| 2-hexanone | 43.1 | S41 | 11.5 | 124.16 | S42 | 327.2 | 306.3 | 0.94 |
| 2-heptanone | 47.1 | S41 | 10.6 | 140.76 | S43 | 317.7 | 326.0 | 1.03 |
| 2-octanone | 51.8 | S44 | 10.3 | 157.45 | S45 | 313.2 | 341.5 | 1.10 |
| 2-decanone | 60.9 | S46 | 9.5 | 190.55 | S47 | 306.6 | 335.8 | 1.10 |
| 2-dodecanone | 71.83 | S34 | 10.3 | 223.87[b] | | 309.7 | | |
| diethyl ether | 27.19 | S34 | 0.76 | 104.74 | S48 | 259.6 | 250.6 | 0.97 |
| dipropyl ether | 35.68 | S34 | −0.9 | 137.68 | S49 | 259.1 | 261.0 | 1.01 |
| dibutyl ether | 44.56 | S34 | −1.8 | 170.45 | S50 | 261.4 | 280.2 | 1.07 |
| dipentyl ether | 54.50 | S34 | −1.9 | 203.40 | S51 | 267.9 | 286.9 | 1.07 |
| dihexyl ether | 64.10 | S34 | −2.3 | | | | | |
| pentane | 26.43 | S34 | | 116.11 | S45 | 206.3 | 220.1 | 1.07 |

TABLE S4 (continued)

| | | | | | | | |
|---|---|---|---|---|---|---|---|
| hexane | 31.56 | S33 | 131.57 | S45 | 221.0 | 230.4 | 1.04 |
| heptane | 36.57 | S33 | 147.45 | S52 | 231.2 | 256.2 | 1.11 |
| octane | 41.49 | S33 | 163.52 | S53 | 238.6 | 266.5 | 1.12 |
| nonane | 46.41 | S33 | 179.69 | S45 | 244.5 | 274.6 | 1.12 |
| decane | 51.38 | S33 | 195.90 | S53 | 249.6 | 282.2 | 1.13 |
| dodecane | 61.51 | S33 | 228.47 | S53 | 258.4 | 289.6 | 1.12 |
| tridecane | 66.43 | S33 | 244.91 | S54 | 261.1 | 299.9 | 1.15 |
| tetradecane | 71.30 | S33 | 261.09 | S53 | 263.6 | 302.8 | 1.15 |
| pentadecane | 76.11 | S33 | 277.66 | S54 | 265.2 | 307.5 | 1.16 |
| hexadecane | 81.38 | S33 | 294.04 | S53 | 268.3 | 305.3 | 1.14 |

[a]values of the physical constants needed to determine internal pressure are listed in Table1 and in references [S55-S57]; [b]estimated value from molar volumes of lower 2-alkanones

# TABLE S5

Excess molar enthalpies, $H_m^E$, volumes, $V_m^E$, and isochoric internal energies, $U_{Vm}^E$, at equimolar composition and 298.15 K, for $n$-alkanoate (1) + $n$-alkane (2) mixtures.

| $n$-$C_n$ | $H_m^E$ /J mol$^{-1}$ | Ref. | $V_m^E$ /cm$^3$ mol$^{-1}$ | Ref. | $U_{Vm}^E$ / J mol$^{-1}$ |
|---|---|---|---|---|---|
| methyl propanoate + $n$-alkane | | | | | |
| 5  | 1191 | S14 | 0.444 | S1  | 1075 |
| 6  | 1345 | S16 | 0.825 | S16 | 1121 |
| 7  | 1425 | S17 | 1.000 | S1  | 1133 |
| 8  | 1535 | S18 | 1.123 | S18 | 1200 |
| 10 | 1685 | S20 | 1.252 | S20 | 1299 |
| 12 | 1831 | S21 | 1.318 | S21 | 1419 |
| 14 | 1930 | S23 | 1.322 | S23 | 1505 |
| 15 | 1966 | S58 | 1.396 | S25 | 1514 |
| 16 | 2019 | S26 | 1.362 | S26 | 1580 |
| 17 | 2065 | S27 | 1.411 | S27 | 2589 |
| methyl butanoate + $n$-alkane | | | | | |
| 5  | 979  | S14 | 0.098 | S1  | 953  |
| 6  | 1100 | S16 | 0.528 | S16 | 957  |
| 7  | 1170 | S17 | 0.744 | S26 | 954  |
| 8  | 1239 | S18 | 0.865 | S18 | 982  |
| 10 | 1351 | S20 | 1.003 | S20 | 1032 |
| 12 | 1483 | S21 | 1.075 | S21 | 1148 |
| 14 | 1577 | S23 | 1.111 | S23 | 1222 |
| 15 | 1626 | S58 | 1.136 | S25 | 1260 |
| 16 | 1661 | S26 | 1.153 | S26 | 1292 |
| 17 | 1690 | S27 | 1.168 | S27 | 1298 |
| methyl pentanoate + $n$-alkane | | | | | |
| 5  | 830  | S124 | $-0.328$ | S15 | 918  |
| 6  | 894  | S16  | 0.235    | S16 | 830  |
| 7  | 968  | S17  | 0.464    | S15 | 833  |
| 8  | 1018 | S18  | 0.617    | S18 | 834  |
| 10 | 1139 | S20  | 0.769    | S20 | 902  |
| 12 | 1240 | S21  | 0.853    | S21 | 974  |
| 14 | 1325 | S23  | 0.913    | S23 | 1033 |
| 16 | 1406 | S26  | 0.966    | S26 | 1096 |
| 17 | 1436 | S27  | 0.9733   | S27 | 1110 |

TABLE S5 (continued)

methyl hexanoate + *n*-alkane

| | | | | | |
|---|---|---|---|---|---|
| 5  | 721  | S14 | −0.465 | S1  | 843  |
| 6  | 782  | S16 | 0.037  | S16 | 772  |
| 7  | 838  | S17 | 0.278  | S1  | 757  |
| 8  | 888  | S18 | 0.442  | S18 | 757  |
| 9  | 936  | S19 | 0.537  | S1  | 774  |
| 10 | 982  | S20 | 0.612  | S20 | 795  |
| 12 | 1081 | S21 | 0.716  | S21 | 860  |
| 13 | 1123 | S22 | 0.755  | S1  | 885  |
| 14 | 1163 | S23 | 0.780  | S23 | 915  |
| 16 | 1233 | S26 | 0.827  | S26 | 970  |
| 17 | 1278 | S27 | 0.829  | S27 | 1003 |

ethyl pentanoate + *n*-alkane

| | | | | | |
|---|---|---|---|---|---|
| 5  | 501  | S5 | −0.344 | S5 | 592  |
| 7  | 762  | S5 | 0.383  | S5 | 651  |
| 9  | 895  | S5 | 0.609  | S5 | 712  |
| 13 | 1118 | S5 | 0.816  | S5 | 862  |
| 15 | 1217 | S5 | 0.874  | S5 | 939  |
| 17 | 1322 | S5 | 0.921  | S5 | 1018 |

ethyl hexanoate + *n*-alkane

| | | | | | |
|---|---|---|---|---|---|
| 5  | 406  | S5 | −0.532 | S5 | 547  |
| 7  | 655  | S5 | 0.222  | S5 | 591  |
| 9  | 778  | S5 | 0.452  | S5 | 642  |
| 13 | 980  | S5 | 0.676  | S5 | 768  |
| 15 | 1086 | S5 | 0.743  | S5 | 850  |
| 17 | 1177 | S5 | 0.791  | S5 | 916  |

propyl ethanoate + *n*-alkane

| | | | | | |
|---|---|---|---|---|---|
| 5  | 1024 | S7 | 0.131 | S7 | 990  |
| 7  | 1199 | S7 | 0.819 | S7 | 969  |
| 9  | 1335 | S7 | 0.964 | S7 | 1145 |
| 13 | 1601 | S7 | 1.144 | S7 | 1239 |
| 15 | 1718 | S7 | 1.167 | S7 | 1344 |
| 17 | 1867 | S7 | 1.211 | S7 | 1466 |

butyl ethanoate + *n*-alkane

| | | | | | |
|---|---|---|---|---|---|
| 5 | 789 | S31 | −0.224 | S31 | 848 |

TABLE S5 (continued)

| | | | | | |
|---|---|---|---|---|---|
| 7  | 980  | S31 | 0.548 | S31 | 821  |
| 9  | 1123 | S31 | 0.754 | S31 | 887  |
| 13 | 1347 | S31 | 0.935 | S31 | 1053 |
| 15 | 1447 | S31 | 0.996 | S31 | 1129 |
| 17 | 1575 | S31 | 1.053 | S31 | 1226 |

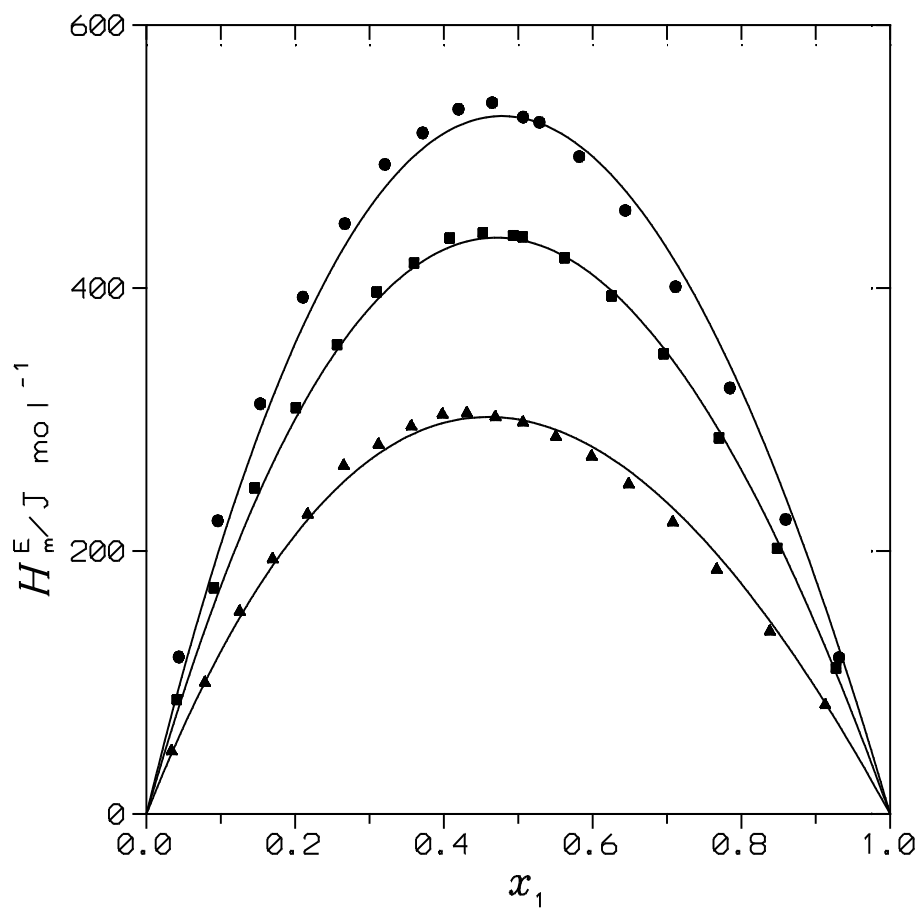

**Figure S1**

$H_m^E$ for $CH_3(CH_2)_{u-1}COOCH_2CH_3$ (1) + heptane (2) systems at 298.15 K. Symbols, experimental results (for source of data, see Table S2): (●), $u = 6$; (■), $u = 7$; (▲), $u = 9$. Lines, Flory results using interaction parameters listed in Table S2.

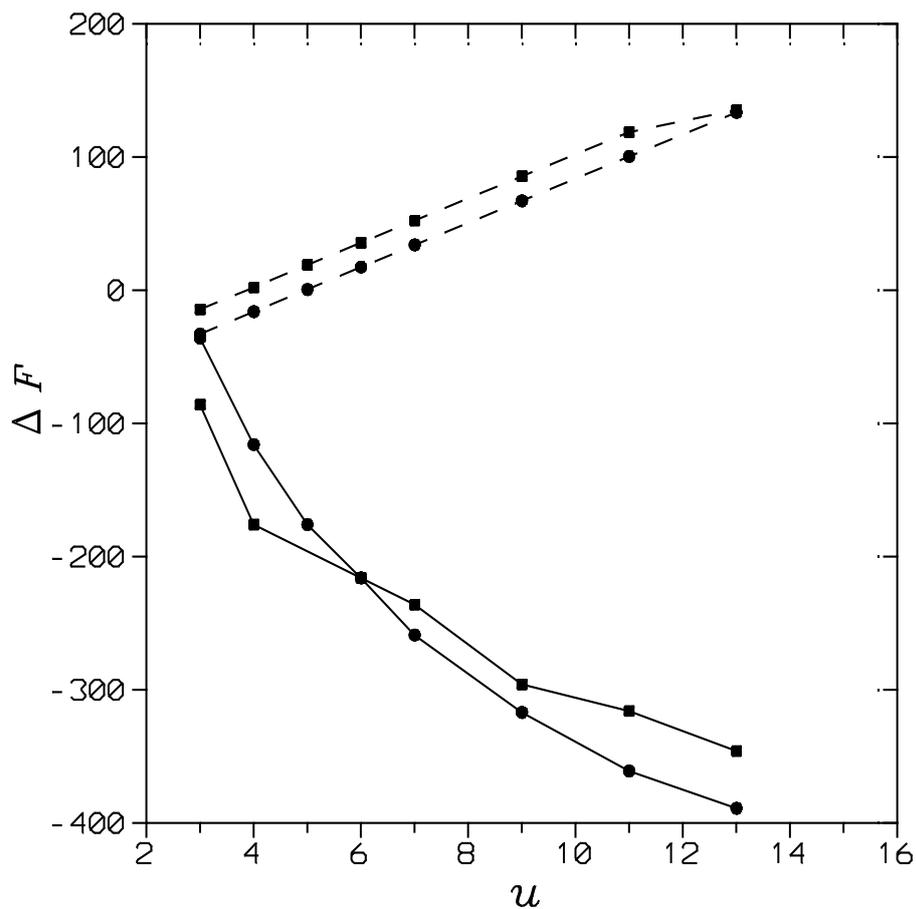

**Figure S2**

Differences between isobaric expansion coefficients or molar volumes of pure compounds in the mixtures $CH_3(CH_2)_{u-1}COO(CH_2)_{v-1}CH_3$ (1) + heptane (2) at 298.15 K. Symbols: (●), $\Delta F / K^{-1} = (\alpha_{p1} - \alpha_{p2})10^3$ ;(■), $\Delta F / cm^3 mol^{-1} = V_{m1} - V_{m2}$. Solid lines, methyl alkanoates, dashed lines ethyl alkanoates. For source of data, see Table S2 and [S55].

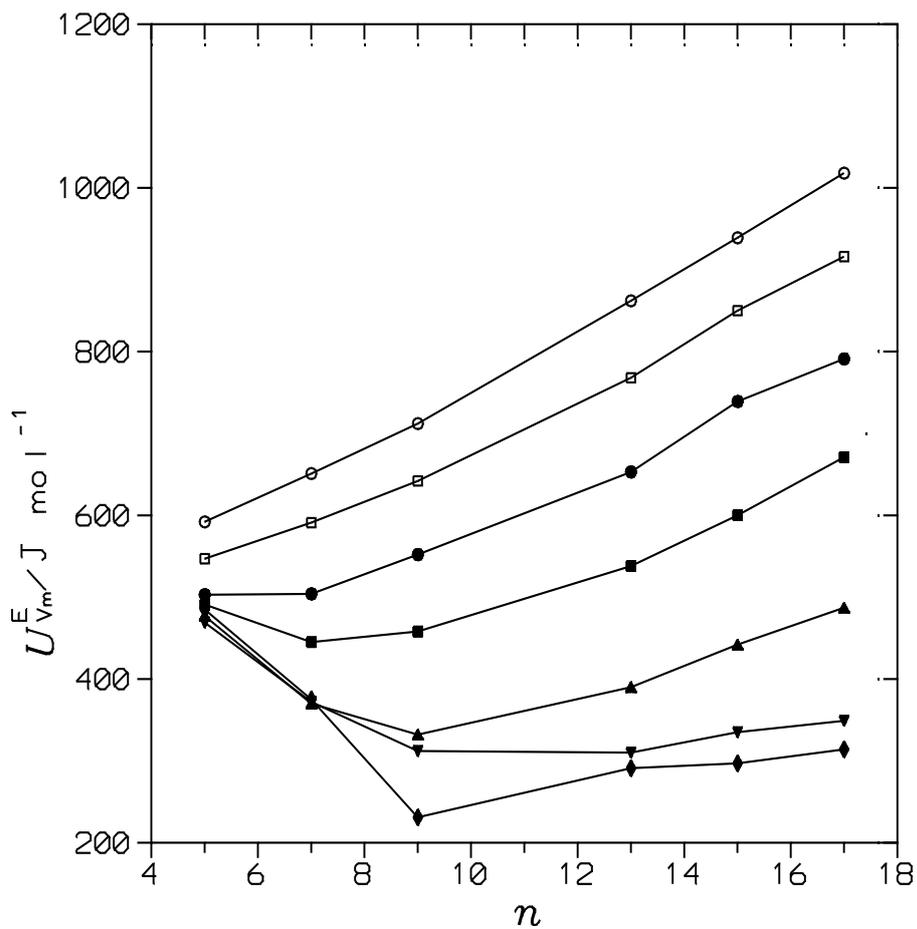

**Figure S3**

$U_{Vm}^E$ of $CH_3(CH_2)_{u-1}COOCH_2CH_3$ (1) + $n$-alkane (2) systems at equimolar composition and 298.15 K and atmospheric pressure vs. $n$, the number of C atoms in the $n$-alkane. Symbols: (○), $u = 4$; (□), $u = 5$; (●), $u = 6$; (■), $u = 7$; (▲), $u = 9$; (▼), $u = 11$; (♦), $u = 13$. Numerical data are listed in Tables S2 and S5.